\documentclass[a4paper,11pt]{article}
\usepackage{amsmath,amssymb,graphicx}
\usepackage{cite}
\usepackage{array,multirow}
\usepackage{multicol,color,longtable}
\usepackage[hyphens,spaces,obeyspaces]{url}
\usepackage{breakurl}
\usepackage{hyperref}
\definecolor{darkred}{rgb}{0.65,0.15,0}
\hypersetup{pdfborder={0 0 0},colorlinks=true,urlcolor=blue,citecolor=blue,linkcolor=darkred,linktocpage=true}

\newcommand{\eprintN}[1]{{\href{http://arxiv.org/abs/#1}{[\texttt{#1 [hep-th]}]}}}
\newcommand{\eprintNT}[1]{{\href{http://arxiv.org/abs/#1}{[\texttt{#1 [math-NT]}]}}}
\newcommand{\doi}[2]{{\hypersetup{urlcolor=darkred}\href{http://dx.doi.org/#2}{#1}\hypersetup{urlcolor=blue}}}

\newcommand{\nn}{\nonumber}

\renewcommand{\Im}{\mathrm{Im}}

\newcommand{\ints}{\mathbb{Z}}

\newcommand{\Li}{\textrm{Li}}
\newcommand{\Lam}{\mathcal{L}}

\newcommand{\esv}{\mathrm{esv}}
\newcommand{\yO}{\;\!\mathsf{y}}
\renewcommand{\mod}{\,\mathrm{mod}\,}

\newenvironment{psmallmatrix}
  {\left(\begin{smallmatrix}}
  {\end{smallmatrix}\right)}

\makeatletter

\@addtoreset{equation}{section} \makeatother

\begin{document}
\setcounter{page}{0}

{\flushright {DCPT-20/01}\\[25mm]}

\begin{center}
{\LARGE \bf Resurgent expansion of Lambert series\\[5mm] and iterated Eisenstein integrals}\\[10mm]

\vspace{8mm}
\normalsize
{\large  Daniele Dorigoni${}^{1}$ and Axel Kleinschmidt${}^{2,3}$}

\vspace{10mm}
${}^1${\it Centre for Particle Theory \& Department of Mathematical Sciences\\
 Durham University, Lower Mountjoy, Stockton Road, Durham DH1 3LE, UK}
\vskip 1 em
${}^2${\it Max-Planck-Institut f\"{u}r Gravitationsphysik (Albert-Einstein-Institut)\\
Am M\"{u}hlenberg 1, DE-14476 Potsdam, Germany}
\vskip 1 em
${}^3${\it International Solvay Institutes\\
ULB-Campus Plaine CP231, BE-1050 Brussels, Belgium}

\vspace{20mm}

\hrule

\vspace{10mm}

\begin{tabular}{p{13cm}}
{\small
We consider special Lambert series as generating functions of divisor sums and determine their complete transseries expansion near rational roots of unity. Our methods also yield new insights into the Laurent expansions and modularity properties of iterated Eisenstein integrals that have recently attracted attention in the context of certain period integrals and string theory scattering amplitudes.
}
\end{tabular}
\vspace{7mm}
\hrule
\end{center}

\thispagestyle{empty}

\newpage

\setcounter{page}{1}
\setcounter{tocdepth}{2}
\tableofcontents

\vspace{5mm}
\hrule
\vspace{5mm}

%%  Sec. 1: Intro  %%
\section{Introduction}\label{Introduction}

The central object of study in this paper is the $q$-series
\begin{align}
 \label{eq:Sabintro}
S_{\alpha,\beta}(q)= \sum_{n,m\geq 1} n^{-\alpha} m^{-\beta} q^{nm}
\end{align}
and we shall be interested in its asymptotic expansion as $q$ approaches a rational root of unity from within the unit disk. For different values of $\alpha$ and $\beta$, the $q$-series $S_{\alpha,\beta}=S_{\beta,\alpha}$ is related to special types of Lambert series that serve as generating functions of divisor sums, to holomorphic Eisenstein series or iterated integrals thereof. 

Our method for analysing the asymptotic expansion makes use of results by Zagier~\cite{ZagierApp} and the transseries completion uses resurgent methods~\cite{Ecalle:1981,delabaere1999resurgent,Dunne:2016jsr}. The completion captures the terms that are exponentially suppressed in the asymptotic expansion and, as we shall see, they are very closely related to the modular properties of the $q$-series. In particular, we shall be interested in something that can be called the `modularity gap'~\cite{Shimomura} which is the failure of the $q$-series to transform as a modular form of definite weight given by $1-\alpha-\beta$ under the $S$-transformation transformation $\tau \to -\frac{1}{\tau}$ where $q=e^{2\pi i \tau}$. 

Depending on the values of $\alpha$ and $\beta$, this modularity gap is given by a Laurent polynomial or by a multi-valued function of $\tau$ whose form depends on a choice of resummation. As the $q$-series is invariant under the $T$-transformation $\tau\to\tau+1$ and since $S$ and $T$ together generate $PSL_2(\ints)$ when acting on $\tau$, we obtain the complete behaviour of the $q$-series $S_{\alpha,\beta}(q)$ under the modular group $PSL_2(\ints)$. As all rational roots of unity are conjugate to $q=1$ under the modular group, one can in principle determine the complete form of $S_{\alpha,\beta}(q)$ expanded around any rational root of unity. However, as this is not necessarily the most convenient way we shall also discuss asymptotic expansions near roots of unity directly.

The case of divisor sums corresponds to $S_{\alpha,0}(q)$ for which the asymptotic perturbative expansion in the limit $q\to 1^{-}$ has already been analysed in~\cite{Knopp,Banerjee}. Our approach gives a concise re-derivation of their result, generalises it to other rational roots of unity and provides the complete transseries expansion. This will be the content of section~\ref{sec:Lambert}. This analysis can also be extended to $q$-Pochhammer symbols as we show in an appendix.

For other special integral choices of $\alpha$ and $\beta$, we make contact with iterated integrals of Eisenstein series of the type that have been recently discussed in the literature~\cite{Brown:2014,Broedel:2015hia,Brown:2017qwo,Brown:2017,Broedel:2018izr,Broedel:2018iwv,Richter:2018hun,Broedel:2019vjc} in connection to periods of moduli spaces of genus-one Riemann surfaces and have found ample applications in string scattering amplitudes~\cite{DHoker:2015gmr,Broedel:2015hia,DHoker:2015wxz,Broedel:2018izr,DHoker:2016mwo}. Our methods can be used in this context to determine the Laurent polynomials of these iterated integrals including the terms that have been hard to obtain in~\cite{Broedel:2015hia} as they are related to integration constants of the differential equations the iterated integrals satisfy~\cite{DHoker:2016mwo,Broedel:2018izr,Mafra:2019ddf,Mafra:2019xms,Gerken:2019cxz}. Again, this analysis is deeply tied in with the modular properties of $S_{\alpha,\beta}(q)$. This connection will be analysed in section~\ref{sec:IterInt}. 

One important aspect in the discussion of iterated integrals is given by the conjectural elliptic single-valued map~\cite{Zerbini:2018sox,Broedel:2018izr,Gerken:2018jrq,Zagier:2019eus,Gerken:2019cxz} that we shall discuss in section~\ref{sec:SV}. Our analysis is restricted to what is known as `depth-one' iterated integrals in the literature and the resulting Laurent polynomials therefore contain only standard Riemann zeta values rather than (single-valued) multiple zeta values.

It would be extremely interesting to understand how to apply the methods explained in the present work to the case of higher depth iterated integrals and multiple polylogarithms (or single-valued version thereof). In particular it is only at higher depth that in string scattering amplitudes we start encountering multiple zeta values in the coefficients of the ``perturbative'' expansion. Furthermore the conjectured elliptic single-valued map, see for example \cite{Broedel:2018izr}, would allow us to go from an open string calculation to a closed string one but its action on the perturbative and non-perturbative amplitudes is far from straightforward. We believe our methods could produce new insights into all these extremely interesting problems.

\subsubsection*{Acknowledgements}
We are grateful to Johannes Broedel, Francesca Ferrari, Jens Funke, Herbert Gangl, Jan Gerken, Erik Panzer, Daniel Persson, Oliver Schlotterer and Federico Zerbini for useful discussions and Oliver Schlotterer for comments on the manuscript. DD would like to thank the Albert Einstein Institute and in particular Hermann Nicolai for the hospitality and support during the various stages of this project.

\section{Lambert series}
\label{sec:Lambert}

In the current section, we begin by studying the general $q$-series $S_{\alpha,\beta}(q)$ defined in~\eqref{eq:Sabintro} when one of the parameters vanishes. The resulting $q$-series will be denoted by
\begin{equation} 
\label{Lambert}
 \Lam_s(q) = S_{s,0}(q) = \sum_{k=1}^\infty k^{-s}\frac{ q^{k}}{1-q^k}\,,
\end{equation}
where $s\in\mathbb{C}$ and $q\in\mathbb{C}$ inside the unit disk $\vert q\vert<1$. We have carried out one of the sums in~\eqref{eq:Sabintro} as it has become geometric and in this way one recognises that~\eqref{Lambert} is a special case of a Lambert series for which we have adopted a more standard notation.

It is straightforward to rewrite this Lambert series in two alternative forms making use of the polylogarithm function
\begin{equation} \label{polylog}
 \Li_s(q) = \sum_{k=1}^\infty \frac{q^k}{k^s} \qquad\qquad s \in \mathbb{C}, \, |q|<1
\end{equation}
and of the divisor function $\sigma_{s}(n) = \sum_{d\vert n} d^{s}$, the sum of the $s^{th}$ power of positive divisors of an integer $n$
through the equations
\begin{equation}\label{Lambert2}
\Lam_s(q) = \sum_{n=1}^\infty \text{Li}_s(q^{n}) =\sum_{m=1}^\infty \sigma_{-s}(m) q^m\,.
\end{equation}
The Lambert series (\ref{Lambert}) can then be understood as the generating function for the divisor function $\sigma_{-s}(n)$.

Making furthermore use of the trivial fact $ \sigma_{s}(m)  = m^{s} \sigma_{-s}(m)$ we can also obtain the relation
\begin{equation}
\Lam_{-s}(q) = \sum_{m=1}^\infty \sigma_{s}(m) q^m = \sum_{m=1}^\infty \sigma_{-s}(m) m^s q^m =  (q\partial_q)^s \Lam_{s}(q)\,,
\end{equation}
so we can consider equation (\ref{Lambert}) for $\mbox{Re}\,s\geq 0 $ and obtain $\mbox{Re}\,s< 0 $ by analytic continuation using the (fractional) derivative operator $(q\partial_q)^s$. For negative integers $s$ this operator should be thought of as an integral operator and this is what will be explored in section~\ref{sec:IterInt}.

\subsection{\texorpdfstring{Asymptotic expansion at $q=1$}{Asymptotic expansion at q=1}}
\label{sec:Asy}

We first want to obtain the asymptotic expansion of (\ref{Lambert}) for $q\to 1^-$, meaning from within the unit disk. To this end, we let $q=e^{-2\pi y}$ with $\mbox{Re}\,y>0$ and consider the asymptotic expansion for $y\to 0^+$. An alternative notation that will be used throughout this paper is the modular parameter $\tau$ defined by $q=e^{2\pi i \tau}$ with $\mbox{Im}\,\tau>0$, which is related to the variable $y$ by $y = - i\tau$. 
By slight abuse of notation we will write interchangeably
\begin{align}
\Lam_s(q) = \Lam_s(y) = \Lam_s(\tau)\qquad \mbox{with}\qquad q=e^{-2\pi y} = e^{2 \pi i \tau}\,.
\end{align}

We are then interested in the expression
\begin{equation}
\label{eq:PolyLog1}
\Lam_s(y) = \sum_{n=1}^\infty \text{Li}_s(e^{-2\pi n y})\,.
\end{equation}

This form for the Lambert series under consideration will be our starting point to obtain an asymptotic expansion for $y\to 0^+$.
We notice in fact that we want to obtain the asymptotic expansion for a series of the form $\sum_{m \geq 0} \phi( (m+a) y)$ if we consider $\phi( y) =\Li_s(e^{-2\pi y})$ and $a=0$, and Zagier has proved a very useful result for this situation~\cite{ZagierApp} which we will now review briefly. 

Assume $\phi(y)$ is a smooth function for $y >0$ with all derivatives of rapid decay at infinity. If $\phi(y)$ has an asymptotic expansion around $y=0$ of the form $\phi(y)\sim \sum_{n\geq 0}  b_n y^n$, then the asymptotic expansion of the function summed over its values at shifted argument has the asymptotic expansion around the origin given by
\begin{align}
\label{eq:Zagexp}
\sum_{m \geq 0} \phi( (m+a) y) \sim \frac{I_\phi}{y} + \sum_{n\geq 0} b_n \zeta(-n,a) y^n
\end{align}
where $a>0$. In this expression the Hurwitz zeta function  $\zeta(-n,a)$ arises from the na\"ive interchange of two infinite sums in 
\begin{align}
\sum_{m\geq 0} \phi((m+a)y) &\underset{\text{na\"ive}}{\sim} \sum_{m\geq 0} \sum_{n \geq 0} b_n (m+a)^n y^n = \sum_{n\geq 0} b_n \left( \sum_{m\geq 0} (m+a)^n\right) y^n
= \sum_{n\geq 0} b_n \zeta(-n,a) y^n\,,
\end{align}
where the $m$-sum is divergent and is to interpreted via analytic continuation of the Hurwitz zeta function. As Zagier has shown the only correction needed in addition to this interchange is given by the `Riemann integral term'
\begin{align}
I_\phi = \int_0^\infty \phi(y) dy\,.
\end{align}
This term arises from interpreting the original sum as an approximation to the Riemann integral for small $y$ with $1/y$ being the length of the integration domain. There are also extensions of~\eqref{eq:Zagexp} when $\phi(y)$ is not $C^\infty$ at the origin but includes terms of the form $y^s \log y$ or $y^{s}$ for $\mbox{Re}\,s> -1$, see~\cite{ZagierApp}, that we shall also use later.

We want to adapt the method just outlined to the case under consideration of equation (\ref{eq:PolyLog1}) 
so first we have to consider $\phi( y) =\Li_s(e^{-2\pi y})$ and compute its Taylor expansion near $y=0$ given by
\begin{equation}\label{eq:Taylor}
\Li_s(e^{-2\pi y}) = (2\pi y)^{s-1} \Gamma(1-s) +\sum_{k=0}^\infty \frac{(-2\pi y)^k}{k!}\zeta(s-k)\,.
\end{equation} 

Applying the method of~\cite{ZagierApp}, the asymptotic expansion of the Lambert series for $y\to0^+$ is then formally given by
\begin{equation}
\Lam_s(y) \sim \frac{I_s}{ y} + \sum_{n=1}^\infty (  2\pi n y) ^{s-1} \Gamma(1-s)+ \sum_{k=0}^\infty \frac{(-1)^k}{k!}\zeta(s-k)\sum_{n=1}^\infty ( 2\pi ny)^k\,,
\end{equation}
where the Riemann term $I_s$ is given by 
\begin{align}
I_s=\int_0^\infty \Li_s(e^{- 2\pi  y}) dy =\frac{\zeta(s+1)}{2\pi}\,,
\end{align}
for $\mbox{Re}\,s>0$, although the final result will be valid for any $s\in\mathbb{C}$.
Clearly the above expression is only formal since two non-convergent sums have been interchanged, but if one interprets $\sum_{n=1}^\infty (2\pi ny)^\alpha$ as its analytic continuation $(2\pi y)^\alpha \zeta(\alpha)$ we obtain 
\begin{equation}
\label{eq:LamAsy}
\Lam_s(y) \sim \frac{\zeta(s+1)}{2\pi y} +\zeta(1-s)\Gamma(1-s)(2\pi y)^{s-1} +\sum_{k=0}^\infty \frac{(-2\pi y)^k}{k!} \zeta(-k)\zeta(s-k)\,,
\end{equation}
where $\sim$ denotes asymptotic in the sense of Poincar\'e and \cite{ZagierApp} shows that the above expansion is then the correct asymptotic expansion of the Lambert series~\eqref{Lambert}.

Note that the Riemann integral term can be understood as the limit $k\to -1$ of the asymptotic series
\begin{equation}
\lim_{k\to -1} \frac{(-2\pi y)^k}{\Gamma(k+1)} \zeta(-k)\zeta(s-k) = \frac{\zeta(s+1)}{2\pi y}\,,
\end{equation}
hence we can rewrite (\ref{eq:LamAsy}) to incorporate the Riemann term into the sum and, after shifting $k\to k-1$, we obtain
\begin{equation}
\Lam_s(y) \sim \zeta(1-s)\Gamma(1-s)(2\pi y)^{s-1} +\sum_{k=0}^\infty \frac{(-2\pi y)^{k-1}}{\Gamma(k)} \zeta(1-k)\zeta(s+1-k)\,,\label{eq:LamAsy2}
\end{equation}
where the $k=0$ term is understood as a limit.
If one makes use of the relation $k \zeta(1-k) = -B_k$ for $k\neq 1$, where $B_k$ is the k$^{th}$ Bernoulli number, we see that equation (\ref{eq:LamAsy2}) is exactly the same asymptotic expansion found in a different way in \cite{Banerjee}, see in particular their Theorem 2.2 after the trivial change $s\to -s$ and having set the authors' variable $x=1$.

Some comments are in order for the cases in which $s\to m \in \mathbb{N}=\{0,1,2,\ldots\}$.
In particular we notice two singular terms in equation (\ref{eq:LamAsy2}), namely $\Gamma(1-s) \zeta(1-s)y^{s-1}$ and the $k=m$ term of the series, however it is fairly simple to see that the sum of these two terms has the finite limit
\begin{align}
&\quad \lim_{s\to m}\left[\Gamma(1-s)\zeta(1-s) (2\pi y)^{s-1} + \frac{(-2\pi y)^{m-1}}{\Gamma(m)}\zeta(1-m)\zeta(1-m+s)\right]\nn\\
&= \left[m\zeta'(1-m) -(\log(2\pi y) -\gamma-\psi(m))\, m\,\zeta(1-m)\right]\frac{(-2\pi y)^{m-1}}{m!}\,,
\end{align}
where $\gamma$ is Euler--Mascheroni constant and $\psi(m)=\Gamma'(m)/\Gamma(m)$ denotes the digamma function.
The case $m=0$ has to be understood as a further limit 
\begin{equation}
\lim_{m\to0} \Big[m\zeta'(1-m) -(\log(2\pi y)-\gamma-\psi(m))\, m\,\zeta(1-m)\Big]\frac{(-2\pi y)^{m-1}}{m!} = \frac{\gamma-\log(2\pi y)}{2\pi y}\,.
\end{equation}

We have then the asymptotic expansions for the Lambert series (\ref{Lambert}) specialised to the case $s=m\in\mathbb{N}^*=\{1,2,\ldots\}$
\begin{align}\label{eq:LamAsyInt}
\Lam_m(y) &\sim  \Big[m\zeta'(1-m) - (\log (2\pi y)- \gamma-\psi(m)) m\,\zeta(1-m)\Big]\frac{(-2\pi y)^{m-1}}{m!} \nn\\
&\hspace{50mm}+\sum_{\substack{k=0, \\ k \neq m}}^\infty \frac{(-2\pi y)^{k-1}}{\Gamma(k)} \zeta(1-k)\zeta(1-k+m)\,.
\end{align}
Note that when $m$ is an odd integer  $\zeta(1-k)\zeta(1-k+m)= 0 $ for $k >m+1$ so (\ref{eq:LamAsyInt}) does actually truncate, leading to a finite asymptotic series.

Similarly for the case $s=0$ we have
\begin{equation}\label{eq:LamAsyZ}
\Lam_0(y)  \sim  \frac{\gamma-\log(2\pi y)}{2\pi y} + \sum_{k=0}^\infty \frac{(-2\pi y)^k}{k!} \zeta(-k)^2\,.
\end{equation}

Both asymptotic series (\ref{eq:LamAsyInt}) and (\ref{eq:LamAsyZ}) match with the results of \cite{Banerjee} obtained following a different method.

\subsection{Exponentially suppressed corrections and modular properties}
\label{sec:LamNP}

We note that the asymptotic expansions (\ref{eq:LamAsy2}), (\ref{eq:LamAsyInt}) and (\ref{eq:LamAsyZ}) can be used to numerically compute (\ref{Lambert}) when $q=e^{-2\pi y} \to 1^{-}$. However, they cannot possibly be accurate when $q\to 0$, the centre of the unit disk. The reason is that (\ref{eq:LamAsy2}), (\ref{eq:LamAsyInt}) and (\ref{eq:LamAsyZ}) are only asymptotic expansions and miss terms of the form $\exp(4\pi^2/\log q) = \exp(-2 \pi /y)$, exponentially suppressed in the limit $y\to 0^+$, i.e. $q\to 1^{-}$.

To obtain the complete transseries representation we start from (\ref{eq:LamAsy2}) for $s$ generic with $\mbox{Re}\,s>0$, the cases $s\in\mathbb{N}$ can be obtained as limits.
We first define $m = \left[ \mbox{Re}\,s\right]$ as the integer part of $ \mbox{Re}\,s$, and split (\ref{eq:LamAsy2}) into a finite sum plus an asymptotic tail
\begin{equation}
\Lam_s(y) \sim \zeta(1-s)\Gamma(1-s)(2\pi y)^{s-1} +\sum_{k=0}^{m+1} \frac{(-2\pi y)^{k-1}}{\Gamma(k)} \zeta(1-k)\zeta(s+1-k) + \Lam_s^{T}(y) \,,\label{eq:LamAsy3}
\end{equation}
with the tail given by
\begin{equation}
\Lam_s^{T}(y)  = \sum_{k=m+2}^{\infty} \frac{(-2\pi y)^{k-1}}{\Gamma(k)} \zeta(1-k)\zeta(s+1-k)\,.
\end{equation}
We rewrite this asymptotic series by making use of Riemann's functional equation and by shifting $k\to k+m+2$
\begin{align}
\Lam_s^{T}(y) = -\frac{y^{s-1}}{\pi} \cos\left( \frac{\pi s}{2}\right) \sum_{k=0}^\infty \left(\frac{y}{2\pi}\right)^{k+m+2-s} &\notag(1+(-1)^{k+m}) \Gamma(k+m+2-s)\\
& \times \zeta(k+m+2) \,\zeta(k+m+2-s)\label{eq:Tail1}\,.
\end{align}

Using the known Dirichlet series
\begin{equation}\label{eq:Diric}
\zeta(k+a)\zeta(k) =\sum_{n=1}^\infty \sigma_{-a}(n) n^{-k}\,,
\end{equation}
we can further simplify the asymptotic tail to
\begin{equation}\label{eq:asyTail}
\Lam_s^{T}(y) = -\frac{y^{s-1}}{\pi} \cos\left( \frac{\pi s}{2}\right) \sum_{n=1}^\infty \sigma_{-s}(n)\sum_{k=0}^\infty \left(\frac{y}{2\pi n}\right)^{k+m+2-s} (1+(-1)^{k+m}) \Gamma(k+m+2-s)\,.
\end{equation}
Using first the Dirichlet series to get rid of the two Riemann zeta is a crucial step\footnote{The present discussion is very similar to \cite{Arutyunov:2016etw,Dorigoni:2019yoq} where it was shown that one can reinterpret asymptotic series with factorially growing coefficients ``dressed'' by other particular combinations of Riemann zeta functions as series with simpler coefficients just evaluated at shifted $y\to y/n$ with $n\in\mathbb{N}$ for which it is easier to evaluate the full non-perturbative completions.} since we have thus obtained a very simple series where the variable $y$ has been shifted to $y\to y/n$ with $n\in\mathbb{N}$ and the sum over $k$ is now amenable to standard Borel-Ecalle resummation~\cite{Ecalle:1981,delabaere1999resurgent} which we will now briefly review.

Given a formal asymptotic series for $y\to 0$
\begin{equation}\label{eq:Formal}
f(y) = \sum_{k=0}^\infty y^{k+\alpha+1}\, c_k\, \Gamma(k+\alpha+1)\,,
\end{equation}
we can consider an auxiliary function, called the Borel transform of $f(y)$, defined by
\begin{align}
B(t) = \sum_{k=0}^\infty t^{k+\alpha} c_k\,.
\end{align}
If the series defining the Borel transform has finite radius of convergence we can make use of the known identity
\begin{align}
 y^{k+\alpha+1} \Gamma(k+\alpha+1) = \int_0^\infty  e^{-t}  \left(y t\right)^{k+\alpha }  y\, dt\,,
\end{align}
to obtain an analytic continuation of the formal asymptotic power series $f$ via Borel resummation given by 
\begin{equation}
\mathcal{S}_\theta\left[ f\right](y) = \int_0^\infty e^{-t} B\left(y t\right) y\,dt = \int_0^{e^{i\theta}\infty} e^{-\frac{t}{y}} B(t) dt\,,\label{eq:LateralDef}
\end{equation}
where $\theta =\mbox{arg}\,y$, provided that the direction of integration does not contain singularities of the Borel transform, i.e. it is not a {\em Stokes direction} for $B(t)$.

We can now consider the direction $\theta$ to be independent from the argument of $y$ and the {\em directional} Borel resummation $\mathcal{S}_\theta\left[ f\right](y)$ defines an analytic function in the wedge $\theta-\frac{\pi}{2}< \mbox{arg}\, y<\theta+\frac{\pi}{2}$ of the complex $y$-plane, with asymptotic expansion for $y\to 0$ given precisely by our starting formal power series (\ref{eq:Formal}). If the Borel transform $B(t)$ has no singularities in the wedge of the Borel $t$-plane $\theta_1<\mbox{arg}\,t<\theta_2$ then $\mathcal{S}_{\theta_2}\left[ f\right](y)$ is the analytic continuation of $\mathcal{S}_{\theta_1}\left[ f\right](y)$ on a wider wedge of the complex $y$-plane since they coincide on the common domain of analyticity.

Let us apply this method to the present formal power series (\ref{eq:asyTail}).
We first define the Borel transform of the series as
\begin{equation}\label{eq:Borel}
B(t) = \sum_{k=0}^\infty t^{k+m+1-s} (1+(-1)^{k+m})= t^{m+1-s} \left(\frac{1}{1-t}+\frac{(-1)^{m}}{1+t}\right)\,,
\end{equation}
and the directional Borel resummation for the tail $\Lam_s^{T}(e^{-y})$ can be written as the Laplace transform of $B(t)$
\begin{equation}\label{eq:DirBorel}
\mathcal{S}_\theta\left[ \Lam_s^{T}\right](y) =  -\frac{y^{s-1}}{\pi} \cos\left( \frac{\pi s}{2}\right) \sum_{n=1}^\infty \sigma_{-s}(n) \int_0^{e^{i\theta} \infty} e^{ - \frac{2 \pi n t}{y}} \,B(t) dt\,,
\end{equation}
where $\theta\to0$, which provides an analytic continuation of the asymptotic tail, valid in the wedge $ \mbox{Re} \,y>0$ of the complex $y$-plane.

The function $B(t)$ has two singular directions in the complex $t$-plane, $\mbox{arg} \,t = 0$ and $\mbox{arg} \,t =\pi$. Let us focus on the $\mbox{arg}\,t=0$ direction, which is the only relevant one for the wedge under consideration $\mbox{Re}\,y>0$, and compute the Stokes automorphism, i.e. the discontinuity across a Stokes direction, using Cauchy's formula
\begin{align}
\lim_{\theta\to 0^+} \Big[\mathcal{S}_\theta\left[ \Lam_s^{T}\right](y) - \mathcal{S}_{-\theta}\left[ \Lam_s^{T}\right](y) \Big] &\notag= \mathcal{S}_+ \left[ \Lam_s^{T}\right](y) -\mathcal{S}_- \left[ \Lam_s^{T}\right](y)\\
&\label{eq:StokesAuto}=  -2 i \,y^{s-1} \cos\left(\frac{\pi s}{2}\right) \sum_{n=1}^\infty \sigma_{-s}(n) e^{-\frac{2\pi n}{y}} \,,
\end{align}
where we have defined the two {\em lateral} resummations $\mathcal{S}_{\pm}$ across the Stokes direction $\theta=0$ via the limit $\lim_{\theta\to 0^+} \mathcal{S}_{\pm\theta} = \mathcal{S}_{\pm}$.

The non-vanishing of this Stokes automorphism means that our resummation (\ref{eq:DirBorel}) would give rise to ambiguities in defining a unique value for the starting asymptotic series (\ref{eq:asyTail}) when $y>0$; furthermore, although our asymptotic tail (\ref{eq:asyTail}) is only a formal object it is nonetheless manifestly real for $y>0$ while neither of the two lateral resummations is. 

To obtain a real and unambiguous resummation for $y>0$ we have to consider an average between the two lateral resummations, usually referred to as median resummation \cite{delabaere1999resurgent}:
\begin{equation}
\mathcal{S}_{\text{med}}\left[ \Lam_s^{T}\right](y)  = \mathcal{S}_{\pm}\left[ \Lam_s^{T}\right](y)\pm i \,y^{s-1} \cos\left( \frac{\pi s}{2}\right) \sum_{n=1}^\infty \sigma_{-s}(n) e^{-\frac{2\pi n}{y}}\,.
\end{equation}
This resummation amounts to having subtracted half of the Stokes automorphism (with sign) from the two lateral resummations $\mathcal{S}_\pm$, in more concrete terms this is equivalent to using a principal value prescription to compute the singular integral $\mathcal{S}_0 \left[ \Lam_s^{T}\right](y)$. The median resummation is clearly real and continuous as $\mbox{arg}\,y \to 0$.

The Stokes constant $\cos( \frac{\pi s}{2})$ fixes the imaginary part of the transseries parameter $\sigma$, i.e. the overall piece-wise constant (jumping only at Stokes directions) in front of the non-perturbative terms, to $\mbox{Im} \,\sigma = \pm i  \cos( \frac{\pi s}{2})$. We will make the assumption~\cite{Dorigoni:2019yoq} that the complete transseries parameter does in fact exponentiate, this means we will work under the hypothesis that
\begin{equation}
\sigma = e^{ \pm i \frac{\pi}{2} ( 1-s) }\,,
\end{equation}
where once more the sign is correlated with the choice of resummation. We have verified the validity of this assumption by numerically evaluating both the $q$-series and its proposed transseries to high numerical precision and, at the end of the derivation, we shall provide further evidences for the correctness of our hypothesis by reproducing some well-known results in certain special cases.

We can then provide the complete transseries expression for the Lambert series:
\begin{align}
\label{eq:PMTS} 
\Lam_s(y) &= \zeta(1-s)\Gamma(1-s)(2\pi y)^{s-1} +\sum_{k=0}^{m+1} \frac{(-2\pi y)^{k-1}}{\Gamma(k)} \zeta(1-k)\zeta(s+1-k) \\
&\quad+\mathcal{S}_{\pm}\left[ \Lam_s^{T}\right](y)+(\mp i y)^{s-1} \sum_{n=1}^\infty \sigma_{-s}(n) e^{-\frac{2\pi n}{y}}\,,\nn
\end{align}
valid in the wedge $\mbox{Re}\,y>0$.

Note that the non-perturbative terms take the exact form of the Lambert series (\ref{Lambert2}) when we replace $y \to 1/y$ or equivalently $\tau \to -1/\tau$.
So for general parameter $s$ we obtain
\begin{align}
\label{eq:TS}
\Lam_s(y)&= \zeta(1-s)\Gamma(1-s)(2\pi y)^{s-1} +\sum_{k=0}^{m+1} \frac{(-2\pi y)^{k-1}}{\Gamma(k)} \zeta(1-k)\zeta(s+1-k) \\
&\quad+\mathcal{S}_{\pm}\left[ \Lam_s^{T}\right](y)+(\mp i y)^{s-1} \Lam_s\left(\frac{1}{y}\right)\,,\nn
\end{align} 
that we verified numerically to agree with (\ref{Lambert}) for different values of $y$ and $s$ and a precision of $10^{-150}$.

If we choose the analytic continuation given by the lateral resummation $\mathcal{S}_-$ we can rewrite equation (\ref{eq:PMTS}) to produce a quasi-modular $S$-transformation replacing $ y = - i \tau$
\begin{align}\label{eq:QuasiMod}
\Lam_s(\tau) - \tau^{s-1} \Lam_s\left(-\frac{1}{\tau}\right) &=  \zeta(1-s)\Gamma(1-s)(-2\pi i \tau)^{s-1} +\sum_{k=0}^{m+1} \frac{(2\pi i \tau)^{k-1}}{\Gamma(k)} \zeta(1-k)\zeta(s+1-k) \nn\\
& \phantom{=}+ \mathcal{S}_{-}\left[ \Lam_s^{T}\right](\tau)\,.
\end{align}
Note that we used the lateral resummation $ \mathcal{S}_{-}$ defined above but we could have used the directional resummation $\mathcal{S}_\theta$ along any direction $-\pi<\theta <0$ since all these analytic continuation coincide on their common domain of analyticity.
The modular weight of $\Lam_s(\tau)$ is $1-s$ and we see that for general $s\in\mathbb{C}$ the right-hand side can be understood as the modularity gap for $\Lam_s(\tau)$, given by a perturbative finite degree Laurent polynomial plus the function $\mathcal{S}_{-}\left[ \Lam_s^{T}\right](\tau)$ which is analytic in $\Im\, \tau>0$ with a branch cut along $\mathbb{R}^-$ as one can easily see from (\ref{eq:DirBorel}).
We have checked that our modularity gap reproduces precisely the same result as~\cite{Shimomura}.

It is worth emphasising that firstly the non-perturbative terms are captured completely by the asymptotic perturbative data and secondly they can be rewritten precisely in terms of the original Lambert series by simply changing $\tau \to -1/\tau$. 

Had we chosen the analytic continuation given by the lateral resummation $\mathcal{S}_+$ we can rewrite equation (\ref{eq:PMTS}) to produce a quasi-modular $(-S)$-transformation
\begin{align}\label{eq:QuasiMod2}
\Lam_s(\tau) - (-\tau)^{s-1} \Lam_s\left(-\frac{1}{\tau}\right) &=  \zeta(1-s)\Gamma(1-s)(-2\pi i \tau)^{s-1} +\sum_{k=0}^{m+1} \frac{(2\pi i \tau)^{k-1}}{\Gamma(k)} \zeta(1-k)\zeta(s+1-k) \nn\\
&\phantom{=}+ \mathcal{S}_{+}\left[ \Lam_s^{T}\right](\tau)\,.
\end{align}
Again we used the lateral resummation $ \mathcal{S}_{+}$ but we could have used the directional resummation $\mathcal{S}_\theta$ along any direction $0<\theta <\pi$ as explained above, and the function $\mathcal{S}_{+}\left[ \Lam_s^{T}\right](\tau)$ we obtain is analytic in $\Im\, \tau>0$ with a branch cut along $\mathbb{R}^+$ as one can easily see from (\ref{eq:DirBorel}).

It is interesting to notice that the two elements $S=\begin{psmallmatrix}0 & -1\\1 & \phantom{-}0\end{psmallmatrix}$ and $-S$ are both in $SL_2(\ints)$ but their action on $\tau$ is the same since they correspond to the same element in $PSL_2(\ints)$, i.e. $S\cdot\tau = -S \cdot \tau = -1/\tau$.
Using the slash operator notation $(f\vert_s \gamma)(\tau) = (c\tau+d)^{-s} f(\frac{a \tau+b}{c\tau+d})$ where $\gamma=\begin{psmallmatrix}a & b\\c & d\end{psmallmatrix}\in SL_2(\ints)$ we see from equations (\ref{eq:QuasiMod}) and (\ref{eq:QuasiMod2}) that in general we have
\begin{equation}
 \Lam_s(\tau) - \Big(\Lam_s \,\Big\vert_{1-s} \,S\Big) (\tau) \neq  \Lam_s(\tau) - \Big(\Lam_s \,\Big\vert_{1-s} \,\,\mbox{$(-S)$}\Big) (\tau)\,,
\end{equation}
this means that $S$ and $-S$ act differently on $\Lam_s(\tau)$ for $s \in \mathbb{C}$ generic. The reason is that for generic $s\in\mathbb{C}$ the modularity gap is not a single-valued function. From equation (\ref{eq:QuasiMod}) we see that an $S$ transformation has automorphy factor $\tau^{s-1}$ while the modularity gap contains the function $ \mathcal{S}_{-}\left[ \Lam_s^{T}\right](\tau)$ which is a multi-valued function on the complex $\tau$ plane. Similarly if we perform the transformation $-S$, the automorphy factor is now $(-\tau)^{s-1}$ while the modularity gap contains the multi-valued function $ \mathcal{S}_{+}\left[ \Lam_s^{T}\right](\tau)$.

To summarise, the two different ways to resum the transseries (\ref{eq:TS}) using the two lateral resummations correspond precisely to the two different actions of $S$ and $-S$ on $\Lam_s(\tau)$ which for generic $s\in\mathbb{C}$ do not project to an action of $PSL_2(\ints)$ on $\Lam_s(\tau)$ because of the multi-valuedness of both the automorphy factor and the modularity gap. We will shortly see that $\Lam_s(\tau)$ will be a genuine quasi-modular form with a proper action of $PSL_2(\ints)$ only when $s$ is an odd integer.

If the parameter $s$ becomes an integer we can simplify the general transseries (\ref{eq:TS}) even further.
Let us suppose that $s=m\in\mathbb{N}$ is an odd integer. In this case the directional Borel transform (\ref{eq:DirBorel}) does actually vanish because of the $\cos( \pi s/2)$ factor, hence the transseries (\ref{eq:TS})
simplifies to
\begin{align}\label{eq:TSOdd}
\Lam_m(y)=&  \left[m\zeta'(1-m) +\delta_{m,1} \log(\sqrt{2\pi y}) \right]\frac{(-2\pi y)^{m-1}}{m!} \\
&\notag+\sum_{\substack{k=0, \\ k \neq m}}^{m+1} \frac{(-2\pi y)^{k-1}}{\Gamma(k)} \zeta(1-k)\zeta(1-k+m) +(-1)^{\frac{m-1}{2}} y^{m-1} \Lam_m\left(\frac{1}{y}\right)\,,
\end{align} 
where we used (\ref{eq:LamAsyInt}). Note that the first two lines reproduce exactly the asymptotic expansion discussed in \cite{Banerjee}, and we have checked numerically that with the addition of this infinitely many exponentially suppressed term (\ref{eq:TSOdd}) coincides with (\ref{Lambert}) within the numerical precision of $10^{-150}$ used. We will shortly show that for $m=1$ the proposed non-perturbative terms are crucial and we can prove that (\ref{eq:TSOdd}) is exact in this case.

Replacing $y=-i\tau$ the above equation gives us the quasi-modularity properties of $\Lam_m(\tau)$ for odd integral $m$:
\begin{align}\label{eq:QMOdd}
\Lam_m(\tau)-\tau^{m-1}\Lam_m\left(-\frac{1}{\tau}\right)=&  \left[m\zeta'(1-m) +\delta_{m,1} \log(\sqrt{-2\pi i  \tau}) \right]\frac{(2\pi i \tau)^{m-1}}{m!} \\
&\notag+\sum_{\substack{k=0, \\ k \neq m}}^{m+1} \frac{(2\pi i \tau)^{k-1}}{\Gamma(k)} \zeta(1-k)\zeta(1-k+m)\,.
\end{align} 
The function $\Lam_m(\tau)$ is quasi-modular with weight $1-m$ and modularity gap given by a Taylor--Laurent polynomial in $\tau$ plus possibly a logarithmic term in the case $m=1$.

A comment is in order at this point. We see that for $\Lam_m(\tau)$ with odd integral $m$ both the asymptotic tail (\ref{eq:Tail1}) and the Stokes automorphism (\ref{eq:StokesAuto}), related to the non-perturbative terms, seem to vanish, however our hypothesis that the transseries parameter exponentiates to $\sigma = e^{ \pm i \frac{\pi}{2} ( 1-m) }$ is now crucial.
This is an example of Cheshire-cat resurgence \cite{Dunne:2016jsr,Kozcaz:2016wvy,Dorigoni:2017smz} for which the non-perturbative terms are still present in transseries expansion for the Lambert series $\Lam_s(\tau)$ in the limit $s\to m$ odd integral despite the vanishing of the asymptotic tail. This analysis is very similar to what we have observed in \cite{Dorigoni:2019yoq} in the context of modular graph functions.

When $s=m\in\mathbb{N}$ is instead an even integer we see that the transseries parameter $(\mp i)^{m-1} = \mp i^{m-1}$ is purely imaginary and its only purpose is to cancel the residue at $t=1$ of the directional Borel transform (\ref{eq:DirBorel}) when $\theta \to 0^{\pm}$. This is equivalent to a Cauchy principal value prescription for the integral (\ref{eq:DirBorel}) when $\theta=0$, hence for $s=m$ even we can write (\ref{eq:TS}) as
\begin{align}
\Lam_m(y)= & \left[m\zeta'(1-m) -(\log(2\pi y) -\gamma-\psi(m))\, m\,\zeta(1-m)\right]\frac{(-2\pi y)^{m-1}}{m!}\nn\\
& +\sum_{\substack{k=0, \\ k \neq m}}^{m+1} \frac{(-2\pi y)^{k-1}}{\Gamma(k)} \zeta(1-k)\zeta(1-k+m)\\
&  -(-1)^{m/2} \frac{y^{m-1}}{\pi}  \sum_{n=1}^\infty \sigma_{-m}(n) \,\mbox{p.v.} \int_0^{ \infty} e^{ - \frac{2 \pi n t}{y}} \,B(t)\,dt\,,\nn
\end{align} 
where the perturbative part for the $m=0$ case has to be understood as a further limit as discussed around equation (\ref{eq:LamAsyZ}).

We can evaluate the principal value integral of the Laplace transform of (\ref{eq:Borel}) in terms of exponential integral functions $\mbox{Ei}$ and obtain
\begin{align}
\Lam_m(y)= & \left[m\zeta'(1-m) -(\log(2\pi y) -\gamma-\psi(m))\, m\,\zeta(1-m)\right]\frac{(-2\pi y)^{m-1}}{m!}\\
& \notag+\sum_{\substack{k=0, \\ k \neq m}}^{m+1} \frac{(-2\pi y)^{k-1}}{\Gamma(k)} \zeta(1-k)\zeta(1-k+m) \\
&\notag -(-1)^{m/2} \frac{y^{m-1}}{\pi}  \sum_{n=1}^\infty \sigma_{-m}(n) \left[e^{\frac{2\pi n}{y}} \mbox{Ei}\left(-\frac{2\pi n}{y}\right)+e^{-\frac{2\pi n}{y}} \mbox{Ei}\left(\frac{2\pi n}{y}\right)\right]\,.
\end{align} 
While we have verified that this formula is numerically correct, it is not the most numerically efficient way of evaluating the Lambert series inside the unit disk. For good numerics it is better to use Borel--Pad\'e approximants applied directly to the tail (\ref{eq:Tail1}) without using the Dirichlet series formula (\ref{eq:Diric}) to rewrite the coefficients. 

\subsection{An application to Eisenstein series}
\label{sec:Eis}

In this section we want to revisit the well-known case of holomorphic Eisenstein series in our formalism to provide further consistency checks that the transseries proposed is indeed the correct one.
 
The generating function for the divisor function (\ref{Lambert2}) is closely related to the holomorphic Eisenstein series
\begin{align}
\label{eq:Eis}
G_{2n}(\tau) &= \sum_{(c,d)\neq(0,0)} \frac{1}{(c+d\tau)^{2n}} = 2 \zeta(2n)+\frac{2(2\pi i)^{2n}}{(2n-1)!} \sum_{m=1}^\infty \sigma_{2n-1}(m) q^m\nn\\
& =2\zeta(2n) +G^0_{2n}(\tau) = 2\zeta(2n)\left( 1+\frac{2}{\zeta(1-2n)} \sum_{m=1}^\infty \sigma_{2n-1}(m) q^m\right)\,,\\
&= 2\zeta(2n)\left( 1+\frac{2}{\zeta(1-2n)}\Lam_{1-2n}(q) \right)\,.\nn
\end{align}
for $q=e^{ 2\pi i \tau}$ and $n\in\mathbb{N}$ with $n\geq 1$. Here, we have defined the $q$-series $G_{2n}^0(\tau)$ that equals $G_{2n}(\tau)$ with the constant term removed.
 
As discussed at the beginning of section \ref{sec:Lambert}, we can obtain the transseries expansion for $ \Lam_{-s}(q)$ by applying the fractional derivative operator $(q\partial_q)^{s}$ to $ \Lam_{s}(q)$ for which we have already computed (\ref{eq:TS}).
In particular for the present discussion, we are interested in the case where $s=m$ is an odd positive integer, so it is fairly simple to apply the standard differential operator $(q\partial_q)^m$ to the transseries (\ref{eq:TSOdd}) to obtain
\begin{align}
\Lam_{-m}(y) = &\label{eq:TSneg}\frac{\zeta(1+m)\Gamma(1+m)}{(2\pi y)^{m+1}} +\frac{\zeta(1-m)}{2\pi y}+\zeta(0)\zeta(-m)+(-1)^{\frac{m+1}{2}} y^{-m-1} \Lam_{-m}\left(\frac{1}{y}\right)\,,
\end{align}
where we notice that the second term is present only for $m=1$ and vanishes otherwise due to $\zeta(1-m)$ being evaluated at a negative even number.

From equation (\ref{eq:Eis}) we can easily obtain the transseries expansion for all the Eisenstein series from the above equation upon setting $m=2n-1$ and $y=-i\tau$.
For example we have
\begin{align}
G_2(\tau) &\label{eq:Eis2}= \frac{\pi^2}{3}\left( 1-24 \Lam_{-1}(\tau)\right) \\
&\notag= \frac{\pi^2}{3} \left[1-24\left(\frac{1}{24}-\frac{i}{4\pi \tau} -\frac{1}{24\tau^2} +\tau^{-2} \Lam_{-1}\left(-\frac{1}{\tau}\right)\right)\right]\,,
\end{align}
where we used (\ref{eq:TSneg}) with $m=1$.
We see that our proposed non-perturbative corrections correctly reproduce the quasi-modularity properties of $G_2$.
We can in fact rewrite $ \Lam_{-1}(\tau)$ back in terms of $G_2(-1/\tau)$ in the above equation to obtain
\begin{equation}
G_2(\tau) = \tau^{-2} G_2\left(-\frac{1}{\tau}\right) +\frac{2\pi i}{ \tau} \,,
\end{equation}
which is a standard result in the theory of modular functions, see for example the classic \cite{Apostol}.

Note that in the present case, using equations (\ref{eq:Poch}), (\ref{eq:Qeta}) and (\ref{eq:Eis2}), we have that the relation $q\partial_q \Lam_1(q) = \Lam_{-1}(q)$ is exactly equivalent to:
\begin{equation}
-4\pi i \frac{d}{d\tau} \eta(\tau) = G_2(\tau)\,,
\end{equation}
yet another well known identity \cite{Apostol}. Using the same method one can obtain similar known differential identities for holomorphic Eisenstein series.

Similarly it is simple to see that for $n>1$ equation (\ref{eq:Eis}) combined with the transseries (\ref{eq:TSneg}) at $m=2n-1$ reduces to
\begin{align}
G_{2n}(\tau) &\label{eq:EisGen}= 2\zeta(2n)\left( 1+\frac{2}{\zeta(1-2n)}\Lam_{1-2n}(\tau)\right)\\
&\notag =2\zeta(2n)\left[ 1+\frac{2}{\zeta(1-2n)} \left(\frac{\zeta(2n)\Gamma(2n)}{(-2\pi i \tau)^{2n}} +\zeta(0)\zeta(1-2n) +(-1)^{n} (-i\tau)^{-2n} \Lam_{1-2n}\!\left(-\frac{1}{\tau}\right)\!\right)\!\right]\\
&\notag = \tau^{-2n}\,2\zeta(2n)\left[ 1+  \frac{2}{\zeta(1-2n)} \Lam_{1-2n}\!\left(-\frac{1}{\tau}\right)\right]=\tau^{-2n}G_{2n}\left(-\frac{1}{\tau}\right)\,,
\end{align}
where we made use of Riemann's functional equation to rewrite the first term in parenthesis and used $\zeta(0)=-1/2$ to cancel the constant term.
The non-perturbative terms we propose are crucial and the above equation manifestly shows the modular properties of the Eisenstein series as modular forms of weight $2n$, i.e. $  G_{2n}(\tau) =\tau^{-2n} G_{2n}(-1/\tau)$ valid for $n>1$.

%Finally one can also check the the proposed exponentially suppressed terms are compatible with the multiplicative properties of the Eisenstein series. In particular since the space of modular forms of weight $2k$ is one-dimensional for $2k \in\{ 4, 6, 8, 10, 14\}$ we must have that
%\begin{equation}
%\label{eq:EisId}E_4^2 =E_8,\,\quad E_{4}E_{6}=E_{{10}},\quad E_{4}E_{{10}}=E_{{14}},\quad E_{6}E_{8}=E_{{14}}\,.
%\end{equation}
%
%In particular we can compute for example $E_4$ and $E_8$ using (\ref{eq:EisGen})
%\begin{align}
%E_4(\tau) &= \tau^{-4} +240 \tau^{-4} \Lam_{-3}(e^\frac{-2\pi i}{\tau})\\
%E_4(\tau)^2&\notag= \tau^{-8} +480 \tau^{-8} \left[\Lam_{-3}(e^{\frac{-2\pi i}{\tau}})+120 \Lam_{-3}(e^{\frac{-2\pi i}{\tau}})^2\right]\\
%E_8(\tau) &\notag= \tau^{-8} +480 \tau^{-8} \Lam_{-7}(e^\frac{-2\pi i}{\tau})\,.
%\end{align}
%
%We can rewrite using (\ref{Lambert2}) the combination $\Lam_{-3}+120\Lam_{-3}^2$ appearing in $E_4^2$ and making use of the identity
%\begin{equation}
%\sigma_7(n) = \sigma_3(n) +120 \sum_{m=1}^{n-1} \sigma_3(m) \sigma_3(n-m)\,,
%\end{equation}
%we can easily show that indeed $E_8 = E_4^2$. Similarly one can show that the other relations in (\ref{eq:EisId}) follow from our transseries expansion accompanied by similar identities for the divisor function.

\subsection{Expansions around other roots of unity}
\label{sec:otherroots}

%\textcolor{blue}{[comment on obtaining this from $SL_2(\ints)$ transformations? Comment on $SL_2(\ints)$ vs. $PSL_2(\ints)$ and choice of contour] Commented on this when we first introduce the two lateral resummations}

The method described by Zagier in \cite{ZagierApp} and discussed in the present paper below equation (\ref{eq:Zagexp}) allows us to extract also the asymptotic expansion of (\ref{Lambert}) for $q$ approaching any rational root of unity from within the unit circle.

Let us consider $q=e^{-2\pi y+2\pi i p/c}$ where $y\in\mathbb{R}^+$ and $p,c\in\mathbb{N}$ co-prime, i.e. $(c,p)=1$. 
From the physics point of view we can interpret this setup as an expansion in a background with non-zero topological angle $\theta = 2 \pi p/c$ and inverse coupling constant $ 1/g = 2\pi y$. In particular an expansion at a different cusp for instantons in string theory in the context of $R^4$ curvature corrections has been considered in~\cite{Bergshoeff:2008qq}. We stress that our final results will only apply to the case when $c$ is a prime number but we postpone imposing this restrictions for the moment.

To obtain an asymptotic expansion for $y\to 0^+$ we make use of the representation for the Lambert series given in (\ref{Lambert2}) in terms of polylogarithms and rewrite the sum into congruence classes modulo $c$
\begin{equation}\label{eq:LamShift}
\Lam_s\left(y- i\,\frac{p}{c}\right) =\sum_{h=1}^c \sum_{n=0}^\infty  \Li_s( e^{-2\pi (n +\tilde{h}) \tilde{y}} \,\theta^{ h p})\,,
\end{equation}
where we defined $\tilde{h}=h/c,\,\tilde{y}= cy$ and $\theta = \exp(2\pi i /c)$. This analysis is similar to the one carried out in~\cite{Dorigoni:2019yoq}.

As discussed at the beginning of section \ref{sec:Asy} in \cite{ZagierApp} an asymptotic expansion is derived for series of the form $\sum_{n\geq0} \phi((n+\tilde{h}) \tilde{y})$ which, for $\phi(\tilde{y}) = \Li_s( e^{-2\pi\tilde{y} } \theta^{ h p})$, is precisely of the form (\ref{eq:LamShift}) just presented.
Similarly to what we have seen above we only need the Taylor expansion for $\Li_s( e^{-2\pi \tilde{y}}  \theta^{ h p})$ near $y=0$. However the case $h=c$ has a slightly different expansion, see equation (\ref{eq:Taylor}), from the $h\neq c$ cases, so we prefer to split the sum over $h$ in (\ref{eq:LamShift}) into the $1\leq h \leq c-1$ sum and the $h=c$ term which gives exactly the same series discussed above (\ref{eq:PolyLog1}) with shifted parameter $y \to \tilde{y} =  cy $.

Hence using \cite{ZagierApp} we have
\begin{equation}\label{eq:AsyRoot}
\Lam_s\left(y- i \,\frac{p}{c}\right)= \Lam_s(cy)+\sum_{k=0}^\infty \frac{(-2\pi c y)^{k-1}}{\Gamma(k)} \sum_{h=1}^{c-1}\zeta\left(1-k,\frac{h}{c}\right) \Li_{s+1-k}\left(e^{2\pi i \frac{ hp}{c}}\right)\,.
\end{equation}

Before embarking on the general analysis, we make a few observations on a special case.
The case $c=2$, $p =1$, i.e. $q\to -1$, is the simplest to discuss because in this instance we can make use of the identities 
\begin{align*}
\zeta\left(k,\frac{1}{2}\right) &=  (2^k-1) \zeta(k)\,,\\
\Li_{s}\left(-1\right) & =( 2^{1 - s}-1)\zeta(s)\,,
\end{align*}
so that equation (\ref{eq:AsyRoot}) becomes
\begin{equation}
\Lam_s\left(y-\frac{i}{2}\right) =2(1+2^{-s}) \Lam_s(2y) - \Lam_s(y)- 2^{1-s}\Lam_s(4y)\,,
\end{equation}
or in terms of $q$ variable
\begin{equation}\label{eq:cequal2}
\Lam_s(-q) =2(1+2^{-s}) \Lam_s(q^2) - \Lam_s(q)- 2^{1-s}\Lam_s(q^4)\,.
\end{equation}
 
Of particular interest is the $s=1$ example discussed above, for which we have the relation 
\begin{equation}
\Lam_1(-q) =3 \Lam_1(q^2) - \Lam_1(q)- \Lam_1(q^4)\,,
\end{equation}
which, due to equations (\ref{eq:Poch})--(\ref{eq:Qeta}), can be rewritten in terms of the Dedekind eta function using $q=e^{2\pi i \tau}$ and it becomes
\begin{equation}
\eta\left(\tau+\frac{1}{2}\right) = e^{ \frac{i \pi}{24}} \frac{\eta(2\tau)^3}{\eta(\tau) \eta(4\tau)}\,,
\end{equation} 
a known identity for this modular function~\cite{Apostol}.

We can obtain similar relations for $\mbox{Re}\, s<0$.
In particular for $s=-m$ with $m\in\mathbb{N}$ we have
\begin{equation}
\Lam_{-m}(-q) =2(1+2^{m}) \Lam_{-m}(q^2) - \Lam_{-m}(q)- 2^{1+m}\Lam_{-m}(q^4)\,.
\end{equation}
which thanks to equations (\ref{eq:Eis2})--(\ref{eq:EisGen}) can be rewritten in terms of the Eisenstein series when the integer $m=2n-1$ is odd:
\begin{equation}\label{eq:GHalfPeriod}
G_{2n}\left(\tau+\frac{1}{2}\right) = (2 + 4^n) G_{2 n}(2 \tau) -G_{2 n}(\tau)  - 4^nG_{2 n}( 4\tau)\,,
\end{equation}
and once more $q=e^{2\pi i \tau}$.
This identity can also be derived from (\ref{Lambert2}) making use of the multiplicative property of the divisor function and equation (\ref{eq:sigmaid}) and it is a special case of a more general identity that we derive below in~\eqref{eq:Hecke}, where we also also explain the relation to Hecke operators.
 
\medskip
 
We return now to the study of the general case (\ref{eq:AsyRoot}). 
Starting from \begin{equation}\label{eq:ExpRoot}
\Lam_s\left(y- i \,\frac{p}{c}\right)= \Lam_s(cy)+\sum_{k=0}^\infty \frac{(-2\pi c y)^{k-1}}{\Gamma(k)} \sum_{h=1}^{c-1}\zeta\left(1-k,\frac{h}{c}\right) \Li_{s+1-k}\left(e^{2\pi i \frac{ hp}{c}}\right)\,,
\end{equation}
first we want to rewrite the polylogarithm and Hurwitz zeta functions in the functionally reflected form.
To this end we make use of 
\begin{equation}
\Li_{s+1-k}\left(e^{2\pi i \frac{ hp}{c}}\right)= \frac{\Gamma(k-s)}{(2\pi)^{k-s}} \left[i e^{-i\frac{\pi}{2} (s+1-k)}\zeta(k-s,\frac{\bar{h}}{c})-i e^{i\frac{\pi}{2} (s+1-k)}\zeta(k-s,1-\frac{\bar{h}}{c}) \right]\,,
\end{equation}
where we defined $\bar{h} \equiv ph \mod c$ and $\bar{h}\in \{1,...,c-1\}$, alternatively $h \equiv p^{-1} \bar{h}\mod c$ with $p^{-1}$ the multiplicative inverse of $p$ modulo $c$.
Similarly we have
\begin{align}
\zeta\left(1-k,\frac{h}{c}\right) &= \frac{2 \Gamma(k)}{(2\pi c)^k} \sum_{l=1}^c \zeta\left(k,\frac{l}{c}\right) \cos\left(\frac{\pi k}{2} - 2\pi \frac{h l}{c}\right)\\
&\notag= \frac{2 \Gamma(k)}{(2\pi c)^k} \sum_{l=1}^c \zeta \left(k,\frac{l}{c}\right) \cos\left(\frac{\pi k}{2} - 2\pi \frac{\bar{h}l p^{-1}}{c}\right)\,.
\end{align}
Hence we can write
\begin{align}
\sum_{h=1}^{c-1}\zeta\left(1-k,\frac{h}{c}\right) \Li_{s+1-k}\left(e^{2\pi i \frac{ hp}{c}}\right) &=\label{eq:reflection}\frac{2\Gamma(k)\Gamma(k-s)}{(2\pi)^{2k-s} c^k} \sum_{\bar{h}=1}^{c-1}\sum_{l=1}^c  \zeta\left(k,\frac{l}{c}\right)\zeta\left(k-s,\frac{\bar{h}}{c}\right) \\
&\quad\notag \times\left[ e^{ 2\pi i \frac{\bar{h} l p^{-1} }{c} }\cos\left(\frac{\pi s}{2}\right)+e^{ -2\pi i \frac{\bar{h} l p^{-1} }{c} }\cos\left(\frac{\pi (s-2k)}{2}\right) \right]\,,
\end{align}
where we rewrote $\zeta(k-s,1-\frac{\bar{h}}{c})$ changing summation variable $\bar{h}\to c-\bar{h}$. 
Note that this expression is manifestly vanishing for $k\geq s+2$ when $s$ is an odd integer.

We then obtain the following truncating perturbative expansion for (\ref{eq:AsyRoot}) when $s$ is an odd integer\footnote{While it is not obvious from this expression, the final perturbative asymptotic piece only contains single Riemann zeta values and no Hurwitz zeta values. This follows from the fact that we can in principle obtain the expansion at any rational root of unity by an $SL_2(\ints)$ transformation of the expansion at $q=1$ which we showed above to contain only Riemann zeta values.} 
\begin{align}
\label{eq:PertRoot}
\Lam^{\text{P}}_s\Big(y- i \,\frac{p}{c}\Big)&=-\frac{( c y )^{s-1}}{\pi}\sum_{k=0}^{s+1}(-1)^k\left( \frac{cy}{ 2\pi }\right)^{k-s} \Gamma(k-s) \\
&\quad\quad\quad\times \Bigg[ \zeta(k)\zeta(k-s) (c^{-s}-c^{-k}+1) \left( \cos\left(\frac{\pi s}{2}\right)+\cos\left(\frac{\pi (s-2k)}{2}\right) \right)  \nn\\
&\hspace{-5mm} +c^{-k}\sum_{\bar{h},l=1}^{c-1}\zeta\!\left(k,\frac{l}{c}\right)\zeta\!\left(k-s,\frac{\bar{h}}{c}\right)\left( e^{ 2\pi i \frac{\bar{h} l p^{-1} }{c} }\cos\left(\frac{\pi s}{2}\right)+e^{ -2\pi i \frac{\bar{h} l p^{-1} }{c} }\cos\!\left(\frac{\pi (s\!-\!2k)}{2}\right) \!\right) \!\Bigg].\nn
\end{align}

The terms $c^{-s}-c^{-k}$ come from the $l=c$ term in the sum, while the $+1$ next to them comes
from the $\Lam_s(cy)$ term in (\ref{eq:ExpRoot}) whose perturbative expansion we have already computed.
For example we have 
\begin{align}
\Lam_3\left(y- \frac{i}{3}\right) &= \frac{\pi^3}{14580 y} +\frac{2i\pi^3}{243} -\frac{\zeta(3)}{2} +\frac{11\pi^3 y}{108} +\frac{y^2 (-4\pi i \pi^3 -243 \zeta(3))}{54} +\frac{\pi^3 y^3}{180}\,,\\
 \Lam_3\left(y-\frac{2i}{3}\right) &\notag= \overline{\Lam_3\left(y- \frac{i}{3}\right)}\,.
\end{align}

As a check we can see that the only singular term in $y$ in (\ref{eq:PertRoot}) comes from the $k=0$ term which can be simplified dramatically to
\begin{equation}
\Lam^{\text{P}}_s\Big(y- i \,\frac{p}{c}\Big)= \frac{\zeta(s+1)}{2\pi c^{s+1} y}\,,
\end{equation}
exactly as already derived in a completely different way in \cite{Knopp}.

We will focus now on the asymptotic tail which exists for $s$ not an odd integer.
Apart from the $k=1$ term that has to be understood as a limit, we can set $k$ to be an integer and we can rewrite the expression (\ref{eq:reflection}) as
\begin{align}
\sum_{h=1}^{c-1}\zeta\left(1-k,\frac{h}{c}\right) \Li_{s+1-k}\left(e^{2\pi i \frac{ hp}{c}}\right) &=\frac{2\Gamma(k)\Gamma(k-s)}{(2\pi)^{2k-s} c^k} \sum_{\bar{h}=1}^{c-1}\sum_{l=1}^c  \zeta\left(k,\frac{l}{c}\right)\zeta\left(k-s,\frac{\bar{h}}{c}\right) \\
&\quad\notag \times\cos\left(\frac{\pi s}{2}\right) \left[ e^{ 2\pi i \frac{\bar{h} l p^{-1} }{c} }+(-1)^k e^{ -2\pi i \frac{\bar{h} l p^{-1} }{c} } \right]\,.
\end{align}

Isolating the $l=c$ term in the above expression we have
\begin{equation}
 \sum_{\bar{h}=1}^{c-1}\zeta\left(k,1\right)\zeta\left(k-s,\frac{\bar{h}}{c}\right) \cos\left(\frac{\pi s}{2}\right) \left[ 1+(-1)^k \right] = \zeta(k)\zeta(k-s) (c^{k-s}-1)(1+(-1)^k)\,,
\end{equation}
hence we are left with studying sums of the form
\begin{equation}
\label{eq:Hurconv}
 \sum_{\bar{h}=1}^{c-1}\sum_{l=1}^{c-1}  \zeta\left(k,\frac{l}{c}\right)\zeta\left(k-s,\frac{\bar{h}}{c}\right)   e^{ \pm 2\pi i \frac{\bar{h} l p^{-1} }{c} }\,.
 \end{equation}
As we show in appendix~\ref{app:HurDir}, such expressions can be evaluated in terms of Dirichlet characters for the finite group $\ints/(c\ints)$ and we assume from now on that $c$ is a prime number.

The final result can be expressed via
\begin{align} 
{\chi}^{\pm}(N) =   e^{ \pm 2\pi i \frac{ p^{-1}N }{c} } \chi_0(N)\,,
\end{align}
where the character $\chi_0(N)$ is equal to $0$ when $N\equiv 0 \mod c$ and $1$ otherwise. In terms of these we have (cf.~\eqref{eq:Dirichletapp})
\begin{align}
&\label{eq:Dirichlet}
\sum_{\bar{h}=1}^{c-1}\sum_{l=1}^{c-1}  \zeta\left(k,\frac{l}{c}\right)\zeta\left(k-s,\frac{\bar{h}}{c}\right)  e^{ \pm 2\pi i \frac{\bar{h} l p^{-1} }{c} }= \sum_{N\geq1} \frac{\sigma_{-s}(N) \,c^{2k-s}}{N^{k-s}} \chi^\pm(N)
\end{align}
The resulting tail is then given by
\begin{align}
\Lam^T_s\Big(y-&i\,\frac{p}{c}\Big)=- \frac{( cy)^{s-1} \cos\left(\frac{\pi s}{2}\right)}{\pi}\sum_{k=s+2}^{\infty}\left( \frac{cy}{ 2\pi }\right)^{k-s} \\
&\notag\times \sum_{N\geq0 } \frac{ \sigma_{-s}(N)}{N^{k-s}} \left[ c^{k-s} \left(\chi^-(N)+ (-1)^k \chi^+(N) \right)+\left(1+ c^{-s}-c^{-s} c^{-(k-s)}\right)(1+(-1)^k)\right]\,,
\end{align}
which is very reminiscent of the asymptotic tail (\ref{eq:asyTail}) found previously. We stress that this formula is only valid for $c$ a prime number.
 
We can apply the same technique of Borel resummation as performed above, noticing that the alternating terms $(-1)^k$ in the asymptotic tail will produce singularities of the Borel transform along the \textit{negative} real axis, hence irrelevant for our discussion of a non-perturbative completion. 
As previously argued we will assume that the transseries parameter does indeed exponentiate and we obtain for the non-perturbative terms 
\begin{align}
\label{eq:NPother}
\Lam^{\text{NP}}_s\left(y-i\,\frac{p}{c}\right)  &= \left(  i c^2 y\right)^{s-1} c^{1-s} \left( \sum_{N=1}^\infty \sigma_{-s}(N) \chi^-(N) e^{-2\pi \frac{N}{c^2 y}}\right)\nn\\
&\quad\quad +\left(  i c y\right)^{s-1}(1+c^{-s})\, \Lam_s\left(\frac{1}{cy}\right)- ( i y)^{s-1} c^{-1} \Lam_{s}\left(\frac{1}{y}\right)\nn\\
& = (  i c y)^{s-1} \Lam_s\left( \frac{1}{c^2 y} + i \,\frac{p^{-1}}{c}\right) \,,
\end{align}
where we have used the simplification~\eqref{eq:charsum} derived in the appendix, and again we have picked the sign for the transseries parameter corresponding to the lateral resummation $\mathcal{S}_-$ for the asymptotic perturbative series (\ref{eq:PertRoot}) as previously discussed in section \ref{sec:LamNP}.

Note that~\eqref{eq:NPother} is exactly the expected transformation for a modular form of weight $1-s$. The modular parameter we started with is $\tau = i y + \frac{p}{c}$ and in the limit $y\to 0^+$ it approaches a rational point on the real line which is conjugate, via an $SL_2(\mathbb{Z})$ transformation, to the cusp at $\tau \to i \infty$.
We just need considering the $SL_2(\mathbb{Z})$ matrix 
\begin{equation}
\gamma = \left(\begin{matrix} -N && -M \\ c && -p\end{matrix}\right)\,,
\end{equation}
with $N,M\in \mathbb{Z}$ such that $ N p + M c =1$, which is possible since $(p,c)=1$. 
With this choice of $\gamma$ we have that 
$\gamma\cdot \tau = \frac{i}{c^2 y} -\frac{N}{c}$ and it is obvious that $N \equiv p^{-1} \mod c$. 
Using once more the slash operator notation we would have that a modular form of weight $1-s$ would transform as 
\begin{align}
\Big(f\vert_{1-s} \gamma \Big)(\tau) = (i c y)^{s-1} f(\gamma\cdot \tau) = (i c y)^{s-1} f\left(\frac{ i}{ c^2 y} - \frac{p^{-1}}{c}\right)
\end{align}
precisely as the above non-perturbative completion.

We also observe that if we consider an average over the non-trivial $c^{th}$ roots of unity
\begin{equation}
\sum_{p=1}^{c-1}\Lam_s\left(y-i \,\frac{p}{c}\right)
\end{equation}
we have that the term $\chi^{-}(N)$ in (\ref{eq:NPother}) simply contributes as $-1$ because of the sum over roots of unity which leads to
\begin{align}\notag
\sum_{p=1}^{c-1}\Lam^{\text{NP}}_s\!\left(y-i \,\frac{p}{c}\right)  = &-c^{1-s}\left( i c^2 y\right)^{s-1} \Lam_s\!\left(\frac{1}{c^2 y}\right) + c(1+c^{-s}) \left( i c y\right)^{s-1}  \Lam_s\!\left( \frac{1}{cy}\right)- (i y)^{s-1}  \Lam_{s}\!\left(\frac{1}{y}\right).
\end{align}
Furthermore it is simple to see that, with the use of (\ref{eq:reflection}), the sum over non-trivial roots of (\ref{eq:ExpRoot}) simplifies dramatically the perturbative expansion (\ref{eq:PertRoot}) reducing it to a simple linear combination of our initial perturbative asymptotic series (\ref{eq:LamAsy})
\begin{align}
\sum_{p=1}^{c-1}\Lam^{\text{P}}_s\left(y-i \,\frac{p}{c}\right) = -c^{1-s} \Lam_s^{\text{P}} \left( c^2 y\right) + c(1+c^{-s}) \Lam_s^{\text{P}} \left( c y\right) - \Lam_s^{\text{P}}(y)\,.
\end{align}

Finally we see that each perturbative term combines with a non-perturbative one allowing us to use our complete transseries form (\ref{eq:TS}), so that for prime $c$ we are left with
\begin{equation}
\sum_{p=0}^{c-1} \Lam_s\left(y-i \,\frac{p}{c}\right)= -c^{1-s} \Lam_s( c^2 y) +c (1+c^{-s})\Lam_s( c y)\,,
\end{equation}
which reduces to the special case (\ref{eq:cequal2}) for $c=2$. This equation can also be derived directly from the $q$-series expansion by using properties of the divisor function for prime $c$.

When specialised again to the case $s=1-2n$ with $n\in\mathbb{N}$ we can use (\ref{eq:Eis2})--(\ref{eq:EisGen}) and the above identity becomes
\begin{equation}
\label{eq:Hecke}
\sum_{p=0}^{c-1} G_{2n}\left(\tau+ \frac{p}{c}\right)  = -c^{2n} G_{2n}(c^2\tau)+ (c+c^{2n}) G_{2n}(c\tau) \,,
\end{equation}
valid for $c$ a prime number and generalization of (\ref{eq:GHalfPeriod}). This identity can be understood by recalling that the holomorphic Eisenstein series are eigenfunctions of the Hecke operators $T_m$ acting on holomorphic modular forms of weight $k$ by~\cite[Chap.~6]{Apostol}
\begin{align}
\label{eq:Heckedef}
(T_m f)(\tau) = m^{k-1}\sum_{d|m} d^{-k} \sum_{p=0}^{d-1} f\left(\frac{m\tau+pd}{d^2}\right)\,.
\end{align}
For the case of $f=G_{2n}$ evaluated at argument $c\tau$ and the Hecke operator $T_c$ with $c$ prime we have
\begin{align}
(T_c G_{2n} )( c\tau) &= c^{2n-1} G_{2n}(c^2\tau) + c^{-1} \sum_{p=0}^{c-1} G_{2n}\left(\tau+ \frac{p}{c}\right) \nn\\
&=\sigma_{2n-1}(c) G_{2n}(c\tau) = (1+c^{2n-1}) G_{2n}(c\tau)\,, 
\end{align}
where the second line uses the known Hecke eigenvalue $\sigma_{2n-1}(c)$ of $G_{2n}$. This last equation is equivalent to~\eqref{eq:Hecke}. The Hecke algebra allows obtaining relations similar to~\eqref{eq:Hecke} for Eisenstein series in the case when $c$ is not prime.

We note that the general Lambert series $\Lam_s$ defined in~\eqref{Lambert2} is an eigenfunction of the Hecke operators as the divisor sum satisfies the requisite property
\begin{align}
\label{eq:Heckenec}
\sum_{d|n,m} d^{-s} \sigma_{-s}\left(\frac{mn}{d^2}\right) = \sigma_{-s}(m) \sigma_{-s}(n)
\end{align}
for any $s$ and $m,n>0$. The constraint on the sum is that $d$ has to be a divisor of both $m$ and $n$, i.e. a divisor of $\gcd(m,n)$. The Hecke eigenvalues are
\begin{align}
T_n \Lam_s = \sigma_{-s}(n) \Lam_s\,,
\end{align}
and our Lambert series is clearly Hecke normalised, i.e. the coefficient $a_1$ in front of the $q^1$ term in the $q$-series expansion (\ref{Lambert2}) is simply $a_1=1$.
This could also be used to obtain expansions of $\Lam_s$ around roots of unity when $c$ is not prime. One does not require exact modularity of $\Lam_s$ for this, the almost modular transformation with weight $k=1-s$ is sufficient.

\section{Generalised iterated Eisenstein integrals}
\label{sec:IterInt}

We now return to the study of the more general $q$-series
\begin{equation}
 \label{eq:Sab}
S_{\alpha,\beta}(q)= \sum_{n,m\geq 1} n^{-\alpha} m^{-\beta} q^{nm}\,,
\end{equation}
for which clearly we have $S_{\alpha,\beta}(q)=S_{\beta,\alpha}(q)$.
This series converges absolutely for all $\alpha,\beta\in\mathbb{C}$ provided $|q|<1$.

Similar to the treatment for Lambert series it is very simple to show the following identities
\begin{align}
\label{eq:SabItInt}
S_{\alpha,\beta}(q) &=\sum_{N\geq1} \frac{\sigma_{\alpha-\beta}(N)}{N^{\alpha}} q^N = \sum_{N\geq1} \frac{\sigma_{\beta-\alpha}(N)}{N^{\beta}} q^N\\
&\notag = \sum_{n\geq 1} n^{-\alpha} \Li_\beta(q^n) =\sum_{m\geq 1} m^{-\beta} \Li_\alpha(q^m) \\
&\notag = (q\partial_q)^{-\alpha} \Lam_{\beta-\alpha}(q) =(q\partial q)^{-\beta} \Lam_{\alpha-\beta}(q)\,,
\end{align}
where again the operator $(q\partial_q)^{-\alpha}$ is to be thought of as a fractional derivative or fractional integral operator, depending on the sign of $\alpha$. (Similarly for $(q\partial_q)^{-\beta}$.)
Furthermore we also have $q\partial_q S_{\alpha,\beta} = S_{\alpha-1,\beta-1}$ which will be useful later on.

Once more we will make use of \cite{ZagierApp} and, as discussed at the beginning of section \ref{sec:Asy}, we change variable $q=e^{-2\pi y}$ and rewrite
\begin{align}
S_{\alpha,\beta}(y) = y^\alpha \sum_{n\geq 1} (n y) ^{-\alpha} \Li_\beta(e^{-2\pi n y}) = y^\alpha \sum_{n \geq 1} \phi_{\alpha,\beta}( n y)\,,
\end{align}
where $\phi_{\alpha,\beta}(y) = y^{-\alpha} \Li_\beta(e^{-2\pi y})$.

Proceeding as we did before we first obtain, using (\ref{eq:Taylor}), the expansion of $\phi_{\alpha,\beta}(y)$ near $y=0$ 
\begin{align}
\phi_{\alpha,\beta}(y) \sim  (2\pi)^{\beta-1} y^{\beta-\alpha-1} \Gamma(1-\beta) +\sum_{k=0}^\infty \frac{(-2\pi)^k y^{k-\alpha}}{k!}\zeta(\beta-k)\,,
\end{align}
while the Riemann term is given by
\begin{align}
I_{\alpha,\beta} = \int_0^\infty \phi_{\alpha,\beta}(y) dy = (2\pi)^{\alpha-1} \Gamma(1-\alpha) \zeta(\beta-\alpha+1) \,.
\end{align}
Note that this integral is convergent only for $\mbox{Re}\,\alpha<1\leq \mbox{Re}\,\beta$ or $\mbox{Re}\,\alpha < \mbox{Re}\,\beta<1$, however the asymptotic expansion we will derive will actually be valid for all $\alpha,\beta\in\mathbb{C}$.  
The reason is that for the integral $I_{\alpha,\beta}$ to be divergent, $\phi_{\alpha,\beta}(y)$ must have non-integrable singularities $y^{-s}$ with $\mbox{Re}\,s\geq1$ at the origin but these singular terms can be treated separately, see \cite{ZagierApp}, and we can view the expression above as the correct analytic continuation valid also outside the domain of convergence of the integral as a function of $\alpha$ and $\beta$.

Proceeding with the method described in \cite{ZagierApp} we have the asymptotic expansion
\begin{align}
\sum_{n\geq 1} \phi_{\alpha,\beta}(n y) &\sim \frac{I_{\alpha,\beta}}{y} + (2\pi)^{\beta-1}  \Gamma(1-\beta) \sum_{n=1}^\infty (n y)^{\beta-\alpha-1} +\sum_{k=0}^\infty \frac{(-2\pi )^k \zeta(\beta-k)}{k!} \sum_{n=1}^\infty (n y)^{k-\alpha}
\end{align}
which we analytically continue to derive the asymptotic expansion for $S_{\alpha,\beta}(q)$ when $q=e^{-2\pi y}$ and $y\to 0^+$:
\begin{align}
\label{eq:AsySab}
S_{\alpha,\beta}(q) &\sim \Gamma(1-\alpha) \zeta(\beta-\alpha+1) (2\pi y)^{\alpha-1}+\Gamma(1-\beta) \zeta(\alpha-\beta+1) (2\pi y)^{\beta-1} \\
&\quad + \sum_{k=0}^\infty \frac{(-2\pi y)^{k}}{k!} \zeta(\alpha-k)\zeta(\beta-k)\,,\nn
\end{align}
which is manifestly symmetric in $\alpha \leftrightarrow \beta$.

Note that for generic $\alpha,\beta \in \mathbb{C}$ the above expression~---~although completely regular~---~is actually a factorially divergent asymptotic series, furthermore when $\alpha$ and/or $\beta$ become integers (\ref{sec:Asy}) or when $\alpha \to \beta$ we have that various terms appear to be singular, however by taking the appropriate limits, as discussed previously in section \ref{sec:Asy}, we can always obtain a perfectly regular, asymptotic power series which only in special circumstances will be truncating as we will shortly see.

\subsection{Iterated Eisenstein integrals}

As we have seen in~\eqref{eq:Eis}, the Lambert series $\Lam_{\beta-\alpha}(q)$ for $\beta-\alpha=1-2n$ with $n\in \mathbb{N}$ is very closely related to holomorphic Eisenstein series without the constant term. More precisely, we have for the holomorphic Eisenstein series $G_{2n}(q)$ without constant term denoted by $G_{2n}^0(q)$ in~\eqref{eq:Eis} that
\begin{align}
\label{eq:LamEis}
G_{2n}^0 (q) =  G_{2n}(q)- 2\zeta(2n)= \frac{4\zeta(2n)}{\zeta(1-2n)} \Lam_{1-2n}(q) \,.
\end{align}
Consider then a fixed $\beta-\alpha=1-2n$ and a non-negative integer $\alpha$. Then~\eqref{eq:SabItInt} shows that $S_{\alpha,\beta}$ leads to $\beta$-fold integral of $G_{2n}^0$, i.e. an iterated Eisenstein integral of the type studied in the literature. To make this connection more precise, we use the notation of~\cite{Broedel:2018izr} and define
\begin{align}
\label{eq:ItInt}
\mathcal{E}_0 (k_1,\ldots, k_r;\tau ) = (-1)^r \int_{0\leq q_1\leq \cdots \leq q_r\leq q} d\!\log q_1\cdots d\!\log q_r \frac{G_{k_1}^0(q_1)}{(2\pi i)^{k_1}}\cdots \frac{G_{k_r}^0(q_r)}{(2\pi i)^{k_r}}\,,
\end{align}
where the $k_i$ are integers in the set $\{0,4,6,8,\ldots\}$ and by convention $G_0^0=-1$. The number of non-zero $k_i$ is called the depth of the iterated integral. In the present work we shall only encounter iterated integrals of depth one, i.e., there is only one non-trivial $G_{k}^0$ but it can be integrated many times. The definition~\eqref{eq:ItInt} excludes the case $k=2$, corresponding to the non-modular Eisenstein series $G_2$ however our analysis can be used to obtain expressions for iterated integrals also of $G_2$. The general $q$-series expansion of the iterated integral~\eqref{eq:ItInt} can be found in~\cite[Eq.~(2.21)]{Broedel:2018izr}.

Our $q$-series are related to the iterated integrals~\eqref{eq:ItInt} by
\begin{equation} 
\mathcal{E}_0(\alpha-\beta+1,0^{\alpha-1}; \tau) = -\frac{2 S_{\alpha,\beta}(q)}{(\alpha-\beta)!}\,,
\end{equation}
where we assume that $\alpha-\beta+1=2n$ and $\alpha$ is a non-negative integer forcing $\beta$ to be an integer as well, while $0^{\alpha-1}$ is a shorthand notation for $\alpha-1$ successive zeros. In this formula we have also taken without loss of generality  $\alpha\geq \beta$. Special cases are
\begin{align}
\label{eq:EisLamspecial}
\mathcal{E}_0(k,0^{-1}; \tau)&=-\frac{2\Lam_{1-k}(\tau)}{(k-1)!}\,,\\
\mathcal{E}_0(k,0^{k-2}; \tau) &\notag = -\frac{2\Lam_{k-1}(\tau)}{(k-1)!}\,,
\end{align}
with $k\geq 2$ even integer.

Note that in the present work we are following the convention for iterated integrals given in \cite{Broedel:2018izr}.
The integrals (\ref{eq:ItInt}) under consideration can be written as linear combinations of powers of $\tau$ and the objects:
 \begin{equation}
\mathcal{G}\left[{\begin{matrix} j_1 & j_2 & ... & j_r \\ k_1 & k_2 & ... & k_r\end{matrix}};\, \tau\right] = 
\int\limits_{\tau}^{i\infty}  \tau_r^{j_r} G_{k_r}(\tau_r) d\tau_r \int\limits_{\tau_r}^{i\infty}  \tau_{r-1}^{j_{r-1}} G_{k_{r-1}}(\tau_{r-1})d\tau_{r-1}\,\,... \int\limits_{\tau_2}^{i\infty}  \tau_1^{j_1} G_{k_1}(\tau_1)d\tau_1\,,\label{eq:Brown}
\end{equation}
where again $k_i$ are even positive integers and $j_i$ are non-negative integers.

The theory of iterated integrals (\ref{eq:Brown}) was developed by Brown in \cite{Brown:2014}.
As thoroughly explained in \cite{Broedel:2018izr} one can easily convert Brown's integrals (\ref{eq:Brown}) to (\ref{eq:ItInt}) however in the present work we will only be working with depth one iterated integrals for which we have 
\begin{equation}
\frac{(2\pi i)^{p+1-k}}{p!}\mathcal{G}\left[{\begin{matrix} p\\ k \end{matrix}}\,;\, \tau\right] = \sum_{a=0}^p \frac{(-1)^a}{(p-a)!} (2\pi i \tau)^{p-a} \mathcal{E}_0(k,0^a;\tau) - \frac{2\zeta(k) }{(2\pi i)^k}\frac{(2\pi i \tau)^{p+1}}{ (p+1)!}\,,
\end{equation}
or the inverse relation:
\begin{align}\label{eq:Brown2}
\mathcal{E}_0(k,0^p;\tau)&= \frac{(2\pi i)^{p+1-k}}{p!} \sum_{a=0}^p(-1)^a \binom{p}{a}\tau^{p-a}\,\mathcal{G}\left[{\begin{matrix} a\\ k \end{matrix}}\,;\, \tau\right] + \frac{2\zeta(k) }{(2\pi i)^k}\frac{(2\pi i \tau)^{p+1}}{ (p+1)!}\\
&\notag = \frac{(2\pi i)^{p+1-k}}{p!} \int_\tau^{i\infty}  (\tau-\tau_1)^p G_{k}(\tau_1) d\tau_1+ \frac{2\zeta(k) }{(2\pi i)^k}\frac{(2\pi i \tau)^{p+1}}{ (p+1)!}\,.
\end{align}

The endpoint divergences at the cusp $\tau\to i\infty$ of the above integrals have to be understood in the sense of tangential-basepoint prescription as described in \cite{Brown:2014}, in practice for the present case this simply amounts to the prescription $\int_\tau^{i\infty}  \tau_1^p d\tau_1= -\tau^{p+1}/(p+1)$.

\subsection{Laurent polynomials of iterated Eisenstein integrals}

We will first be interested in iterated integrals $\mathcal{E}_0(k,0^{k-2-\ell}; \tau)$ which means $\alpha = k-\ell-1$ and $\beta = -\ell$ with $k\geq2$ even and $\ell\in\mathbb{Z}$, i.e.
\begin{equation}\label{eq:EklS}
\mathcal{E}_0(k,0^{k-2-\ell}; \tau) = -\frac{2 S_{k-\ell-1,-\ell}(q)}{(k-1)!}\,.
\end{equation}
Since $k$ is even $\alpha$ and $\beta$ have opposite parity.
The cases~\eqref{eq:EisLamspecial} correspond to two special values for $\ell$, namely $\ell=0$   and $\ell=k-1$. The  case $\ell=0$ corresponds precisely to the quasi-modular objects $\Lam_{k-1}(\tau)$ studied above in section~\ref{sec:Lambert}, while $\ell=k-1$ corresponds directly to the modular Eisenstein series $G^0_k(\tau)$, see~\eqref{eq:LamEis}.\footnote{Our formulas will also provide generalisations of~\eqref{eq:ItInt} since in principle  the parameters $\alpha$ and $\beta$ need not be integers so that a direct interpretation of~\eqref{eq:SabItInt} as a bona-fide iterated integral is unavailable.
}

When $\alpha $ and $\beta$ have opposite parity and are related to $k$ as above, the asymptotic series~\eqref{eq:AsySab} actually truncates after $n\geq \alpha+1$. Note furthermore that the first two terms can be singular for $\alpha,\beta$ tending to integers, however, in this limit they are precisely compensated by the terms with the same powers of $y$ coming from the series, i.e. the $n=\alpha-1$ and $n=\beta-1$ terms, which are also singular to produce a finite result.

For example, we obtain the following asymptotic expansions
\begin{align}
\label{eq:Ex400}
 \mathcal{E}_0(4,0,0; \tau) &= -\frac{2}{3!} S_{3,0}(q) =-\frac{2}{3!}  \Lam_3(q)\sim -\frac{2}{3!}\left( \frac{\pi^3}{180y}-\frac{\zeta(3)}{2}+\frac{\pi^3 y}{36} -\frac{\zeta(3) y^2}{2} +\frac{\pi^3 y^3}{180}\right)\,,\nn\\
  \mathcal{E}_0(4,0; \tau) &= -\frac{2}{3!} S_{2,-1}(q) =-\frac{2}{3!} (q\partial_q)  \Lam_3(q) \nn\\
  &\sim  -\frac{2}{3!\times (-2\pi)}\left( -\frac{\pi^3}{180 y^2}+\frac{\pi^3}{36} -\zeta(3) y +\frac{\pi^3 y^2}{60}\right)\,,\\
   \mathcal{E}_0(4,0,0,0; \tau) &= -\frac{2}{3!} S_{4,1}(q) =-\frac{2}{3!} (q\partial_q)^{-1}  \Lam_3(q) \nn\\
   & \sim -\frac{2\times (-2\pi)}{3!} \left(\frac{\pi^3 \log(2\pi y)}{180} - \frac{\zeta'(4)}{2\pi}-\frac{\zeta(3) y}{2}+\frac{\pi^3 y^2}{72} -\frac{\zeta(3) y^3}{6} +\frac{\pi^3 y^4}{720} \right)\,,\nn\\
   \mathcal{E}_0(4,0,0,0,0; \tau) &= -\frac{2}{3!} S_{5,2}(q) =-\frac{2}{3!} (q\partial_q)^{-2}  \Lam_3(q) \nn\\
   & \sim -\frac{2\times (-2\pi)^2}{3!} \bigg(\frac{\pi^3 y \log(2\pi y)}{180} +\frac{\zeta(5)}{24} - \frac{\pi^3 y}{180} - \frac{\zeta'(4)}{2\pi}y
   -\frac{\zeta(3) y^2}{4}+\frac{\pi^3 y^3}{216}\nn\\
   &\hspace{40mm} -\frac{\zeta(3) y^4}{24} +\frac{\pi^3 y^5}{3600} \bigg)\,.\nn
\end{align}
We will refer to these as (holomorphic) Laurent polynomials of iterated integrals and the particular cases above have already been given in the literature~\cite{Brown:2017qwo,Brown:2017,Broedel:2018izr,Zerbini:2018sox}. The novelty of our approach is reflected for example in the constant $\zeta'(4)$ term present in the $\mathcal{E}_0(4,0,0,0; \tau)$ expansion, while everything else could have been derived either by differentiation or integration from the Lambert series $\Lam_3(q)$ studied above. This term is an integration constant that is difficult to determine in the approach of~\cite{Broedel:2018izr} which relies on the Cauchy--Riemann equation satisfied by the iterated integral. 

An example of such a Cauchy--Riemann equation following from the definition~\eqref{eq:ItInt} is
\begin{align}
\partial_\tau \mathcal{E}_0(4,0,0,0;\tau) = 2\pi i  \mathcal{E}_0(4,0,0;\tau)
\end{align}
and integrating this directly as a holomorphic function requires fixing one integration constant which can be fixed by studying the modular behaviour.\footnote{Also, Enriquez' method to infer such integration constants from B-elliptic multiple zeta values does not apply here (cf. appendix of~\cite{Broedel:2018izr}) since $\mathcal{E}_0(4,0,0,0;\tau)$ cannot be realised as an elliptic multiple zeta value.}
Here it follows directly from a limit of our general formula~\eqref{eq:AsySab}. Had we studied $\mathcal{E}_0(6,0^5; \tau)$ we would have found $\zeta'(6)$ appearing and so on. The particular coefficient $\zeta'(4)$ plays a r\^ole in the one-loop four-point amplitude of the open superstring~\cite[Eq.~(4.22)]{Hohenegger:2017kqy}. It is easy to generate the integration constants in  $\mathcal{E}_0(4,0^{2+\ell};\tau) = -\tfrac{2}{3!}(q\partial_q)^{-\ell}  \Lam_3(q)$ for any $\ell>0$ and the cases $\ell=1$ and $\ell=2$ are shown in~\eqref{eq:Ex400}. Doing this one finds that the transcendentality of the integration constant is $3+\ell$ as can be seen from the general formula~\eqref{eq:AsySab}.

\subsection{Transseries completion and modular properties}

From the asymptotic expansion (\ref{eq:AsySab}) it is simple to derive the non-perturbative corrections following the same line of reasoning as the one applied to the Lambert series in section~\ref{sec:LamNP}. Using the functional equation for the Riemann zeta functions we arrive at
\begin{align}\label{eq:SabT}
S_{\alpha,\beta}^{T}(q) = -\frac{(2\pi)^\beta y^{\alpha-1}}{\pi}   \sum_{n=1}^\infty \sigma_{\beta-\alpha}(n) \sum_{k=0}^\infty &\left(\frac{y}{2\pi n}\right)^{k+1-\alpha} \frac{\Gamma(k+1-\alpha)\Gamma(k+1-\beta)}{\Gamma(k+1)}\\
&\notag \left[\cos\left(\frac{\pi(\alpha+\beta)}{2}\right) -(-1)^k \cos\left(\frac{\pi (\alpha-\beta)}{2}\right) \right]\,,
\end{align}
which for $\beta=0$ and $\alpha= s$ reduces precisely to the Lambert case (\ref{eq:asyTail}).
We stress again that for $\alpha, \beta$ integers of opposite parity this tail does actually truncate as is manifest from the cosines.

One can define the Borel transform in this case as 
\begin{align}
B(t) &= \sum_{k=0}^\infty t^{k-\alpha} \frac{\Gamma(k+1-\beta)}{\Gamma(k+1)} \left[\cos\left(\frac{\pi(\alpha+\beta)}{2}\right) -(-1)^k \cos\left(\frac{\pi(\alpha-\beta)}{2}\right) \right]\\
&\notag=  \Gamma(1-\beta)t^{-\alpha} \left[\cos\left(\frac{\pi(\alpha+\beta)}{2}\right)(1-t)^{\beta-1} -\cos\left(\frac{\pi(\alpha-\beta)}{2}\right)(1+t)^{\beta-1}  \right]\,,
\end{align}
which clearly has two singular directions for $\mbox{arg} \,t=0$ and $\mbox{arg}\,t=\pi$.
Note that we chose in here a slightly asymmetric form in $\alpha \leftrightarrow \beta$ only to obtain a simpler Borel transform, we could have insisted in using expressions symmetric in $\alpha \leftrightarrow \beta$ and the Borel transform would have become a hypergeometric function $_2F_1$ without changing anything important in what follows.

From this Borel transform it is simple to compute
\begin{align}
&\notag\lim_{\theta\to 0^+} \mathcal{S}_\theta\left[ S_{\alpha,\beta}^{T}\right](y) - \mathcal{S}_{-\theta}\left[ S_{\alpha,\beta}^{T}\right](y)=\mathcal{S}_+\left[ S_{\alpha,\beta}^{T}\right](y)-\mathcal{S}_-\left[ S_{\alpha,\beta}^{T}\right](y) \\
& = \sum_{n=1}^\infty\left[-2 i  \cos\left( \frac{\pi(\alpha+\beta)}{2}\right)\right] (2\pi)^\beta y^{\alpha-1} \sigma_{\beta-\alpha}(n) \,e^{-\frac{2\pi n}{y}}
\, U\left(\beta,1+\beta-\alpha;\frac{2\pi n}{y}\right)\,,
\end{align}
where $U$ denotes the confluent hypergeometric function.

Assuming as we did above that the transseries parameter $i \cos( \pi(\alpha+\beta)/2)$ exponentiates to $\sigma = \exp(\pm i \frac{\pi}{2} (1-\alpha-\beta))$ we obtain that the non-perturbative completion of the asymptotic series (\ref{eq:AsySab}) becomes
\begin{align}
S_{\alpha,\beta}^{\text{NP}}(y)&=(\mp i y)^{\alpha+\beta-1}\sum_{n=1}^\infty \frac{\sigma_{\beta-\alpha}(n)}{n^\beta}  e^{-\frac{2\pi n}{y}} \left(\frac{2\pi n}{y}\right)^\beta \,U\left(\beta,1+\beta-\alpha;\frac{2\pi n}{y}\right)\\
&\notag =(\mp i y)^{\alpha+\beta-1}\sum_{n=1}^\infty \frac{\sigma_{\beta-\alpha}(n)}{n^\beta}  e^{-\frac{2\pi n}{y}}\,_2F_0\left(\alpha,\beta;-\frac{y}{2\pi n}\right)\,,
\end{align} 
where the sign is according to the direction we choose to resum the Borel transform of (\ref{eq:AsySab}) and in the second expression we rewrote $U(\beta,1+\beta-\alpha;z) =z^{-\beta}\,_2F_0(\alpha,\beta;-z^{-1})$ to make the symmetry $\alpha \leftrightarrow \beta$ manifest again.
In particular notice that for the case $\alpha,\beta$ integers of opposite parity the asymptotic tail vanishes and there is no Laplace integral to be performed, $\alpha+\beta-1$ becomes an even integer and hence the sign does not matter so that, passing to the variable $\tau = i y$ we have
\begin{equation}\label{eq:SabTS}
S_{\alpha,\beta}(\tau)=S^{\text{P}}_{\alpha,\beta}(\tau)+ \tau^{\alpha+\beta-1} \sum_{n=1}^\infty \frac{\sigma_{\beta-\alpha}(n)}{n^\beta}  e^{-\frac{2\pi n i }{\tau}}\,_2F_0\left(\alpha,\beta;\frac{i\tau}{2\pi n}\right)\,,
\end{equation} 
where $S^{\text{P}}_{\alpha,\beta}(\tau)$ is a Laurent polynomial (plus possibly logarithmic terms) obtained by (\ref{eq:AsySab}).

For $\alpha,\beta$ generic we have that the perturbative series does not truncate and are left with
\begin{align}\label{eq:SabTSGen}
S_{\alpha,\beta}(\tau)&=\Gamma(1-\alpha) \zeta(\beta-\alpha+1) (-2\pi i\tau)^{\alpha-1}+\Gamma(1-\beta) \zeta(\alpha-\beta+1) (- 2\pi i\tau)^{\beta-1} \\
&\quad +\mathcal{S}_-\left[ S^T_{\alpha,\beta}\right](\tau)
\notag+ \tau^{\alpha+\beta-1} \sum_{n=1}^\infty \frac{\sigma_{\beta-\alpha}(n)}{n^\beta}  e^{-\frac{2\pi n i }{\tau}}\,_2F_0\left(\alpha,\beta;\frac{i\tau}{2\pi n}\right)\,,
\end{align} 
where $\mathcal{S}_-\left[ S^T_{\alpha,\beta}\right](\tau)$ denotes the lateral Borel resummation of the asymptotic (non-truncating) series (\ref{eq:SabT}), i.e. 
\begin{align}
&
\label{eq:SabFullGen} \mathcal{S}_-\left[ S^T_{\alpha,\beta}\right](\tau) =  -\frac{(2\pi)^\beta (-i \tau)^{\alpha-1}}{\pi}   \sum_{n=1}^\infty \sigma_{\beta-\alpha}(n)\Gamma(1-\beta) \\ 
&\notag\times \int_0^{\infty e^{i\phi}} e^{ - \frac{ 2\pi n i t}{\tau}} t^{-\alpha}  \left[\cos\left(\frac{ \pi(\alpha+\beta)}{2}\right)(1-t)^{\beta-1} - \cos\left(\frac{\pi(\alpha-\beta)}{2}\right)(1+t)^{\beta-1}  \right]\,dt\,,
\end{align}
with $-\pi<\phi<0$.

Note that this integral is not necessarily convergent near $t\sim 0$, but only because we decided to resum via Borel transform the complete asymptotic series (\ref{eq:SabT}), while to have a convergent expression we should have first split the asymptotic series (\ref{eq:SabT}) into a finite order polynomial by keeping $ 0\leq k\leq \left[ \mbox{Re}\,\alpha\right]$, with $\left[\mbox{Re}\, \alpha\right]$ denoting the integer part of $\mbox{Re}\,\alpha$, and an asymptotic tail $k > \left[\mbox{Re}\, \alpha\right]$ that we can Borel resum with a genuine convergent integral.
However for simplicity we prefer to present (\ref{eq:SabFullGen}) and interpret it as analytic continuation in $\alpha$ and this will produce the same results.

We can rewrite both in (\ref{eq:SabTS}) and (\ref{eq:SabTSGen}) the $_2F_0$ in its (asymptotic) Gauss series form to obtain the suggestive
\begin{align}\label{eq:SabSDual}
S_{\alpha,\beta}(\tau)=&S^{\text{P}}_{\alpha,\beta}(\tau)+\sum_{m=0}^\infty \frac{(\alpha)_m(\beta)_m}{m!} \frac{ \tau^{\alpha+\beta+m-1}}{(-2\pi i)^m} S_{\alpha+m,\beta+m}\left(-\frac{1}{\tau}\right)\,,
\end{align} 
where we denoted schematically with $S^{\text{P}}_{\alpha,\beta}(\tau)$ either the truncating perturbative expansion appearing in (\ref{eq:SabTS}), or the perturbative Borel resummed expansion appearing in the general case (\ref{eq:SabTSGen}).
This expression suggests that the functions $S_{\alpha,\beta}(\tau)$ transform as a vector-valued quasi-modular form, with weight $1-\alpha-\beta$ and modularity gap given by $\tau^{1-\alpha-\beta}S^{\text{P}}_{\alpha,\beta}(\tau)$ where the perturbative part is intended as above.

We want to stress again two important points: firstly that the non-perturbative corrections are completely encoded into the asymptotic perturbative data; and secondly that the non-perturbative terms can be written precisely as a (possibly infinite) linear combination of our original functions $S_{\alpha',\beta'}(-1/\tau)$ evaluated at $S$-dual modular parameter.
This phenomenon that the non-perturbative corrections provide exactly the $S$-dual vector-valued modular completion of the asymptotic power series is closely reminiscent of the analysis carried out in \cite{Cheng:2018vpl} in the context of $3$-dimensional $\mathcal{N}=2$ Chern--Simons theories.

Going back to the iterated integrals $\mathcal{E}_0(k,0^{k-2-\ell}; \tau)$ we need to consider $S_{k-\ell-1,-\ell}(\tau)$.
In particular we have two special cases $\ell=0$, corresponding to $\mathcal{E}_0(k,0^{k-2}; \tau)$ and $\ell=k-1$ corresponding to $\mathcal{E}_0(k,0^{-1}; \tau)$.
The key fact is that in both cases the $_2F_0$ in~\eqref{eq:SabTS} reduces to $1$.
Let us focus on the $\ell=0$ case first which give us
\begin{align}
\mathcal{E}_0(k,0^{k-2}; \tau) &\label{eq:E0Lagom}= -\frac{2}{(k-1)!}\left[S^{\text{P}}_{k-1,0}(\tau)+ \tau^{k-2} \sum_{n=1}^\infty \sigma_{1-k}(n)  e^{-\frac{2\pi n i }{\tau}} \right]\\
&\notag =-\frac{2}{(k-1)!} S^{\text{P}}_{k-1,0}(\tau) + \tau^{k-2}\mathcal{E}_0\left(k,0^{k-2}; -\frac{1}{\tau}\right)\,,
\end{align}
where $S^{\text{P}}_{k-1,0}(\tau)$ is precisely the Laurent polynomial obtained for the original Lambert series (\ref{eq:LamAsy3}) for $m = k-1$ odd.

The above equation tells us that $\mathcal{E}_0(k,0^{k-2}; \tau)$ is a quasi-modular form of weight $2-k$ with `modularity gap' given by $ \tau^{2-k}\frac{2}{(k-1)!} S^{\text{P}}_{k-1,0}(\tau)$.
We can derive the same expression starting from Brown's version of the iterated integrals (\ref{eq:Brown2}).
Changing integration variables $\tau_1\to -1/\tau_1$ and using the modularity properties of $G_k(\tau)$ we can easily derive
\begin{equation}
\mathcal{E}_0(k,0^{k-2}; \tau) =  \frac{ r_k(\tau)}{(2\pi i) (k-2)!} +\frac{2\zeta(k) (\tau^{-1} +\tau^{k-1})}{(2\pi i)(k-1)!} + \tau^{k-2}\mathcal{E}_0\left(k,0^{k-2}; -\frac{1}{\tau}\right)\,,
\end{equation}
where $ r_k(\tau)$ is precisely the period polynomial \cite{Zagier:1991} of $G_k^0$ (using the tangential-basepoint regularization as mentioned above):
\begin{align}
 r_k(\tau) &= \int_0^{i\infty} (\tau-\tau_1)^{k-2} G_k^0(\tau_1) d\tau_1\\
 &\notag=  -4 \frac{\zeta(k)}{\zeta(1-k)} \sum_{n=0}^{k-2} \frac{(k-2)!}{(k-2-n)!}  \frac{\zeta(n+1) \zeta(n+2-k)}{(2 \pi i)^{n + 1}} \tau^{k-2-n}\,,
\end{align}
where in this expression we have already explicitly evaluated the L-series associated to $G_k^0$ using (\ref{eq:Eis}).

We can then rewrite the Laurent polynomial discussed above
\begin{equation}\label{eq:cocycle}
-\frac{2}{(k-1)!} S^{\text{P}}_{k-1,0}(\tau) = \frac{ r_k(\tau)}{(2\pi i) (k-2)!} +\frac{2\zeta(k) (\tau^{-1} +\tau^{k-1})}{(2\pi i)(k-1)!}\,,
\end{equation} 
which is then guaranteed \cite{Zagier:1991,Brown:2014} to satisfy the cocycle conditions
\begin{equation}
S^{\text{P}}_{k-1,0}\,\Big\vert_{2-k} (1+S)  = 0\,,\qquad S^{\text{P}}_{k-1,0}\,\Big\vert_{2-k} (1+U+U^2)  = 0\,,
\end{equation}
where $U = TS = \begin{psmallmatrix}1 & -1\\1 & \phantom{-}0\end{psmallmatrix}$ and the notation means the sum of the actions of the $SL(2,\ints)$ group elements.

For the $\ell=k-1$ case we have $\mathcal{E}_0(k,0^{-1}; \tau) = -\frac{G^0_k(q)}{(2\pi i)^k}$, where 
\begin{equation}
G^0_k(q) = \frac{(2\pi i)^k}{(k-1)!} \sum_{n\geq1} \sigma_{k-1}(n)q^n =  \frac{(2\pi i)^k}{(k-1)!} \Lam_{1-k}(q)
\end{equation}
is the standard holomorphic Eisenstein series without constant term.
We can specialise equation (\ref{eq:SabTS}) to the case $\ell=k-1$ and obtain
\begin{align}
\mathcal{E}_0(k,0^{-1}; \tau) &= -\frac{2}{(k-1)!}\left[S^{\text{P}}_{0,1-k}(q)+ \tau^{-k} \sum_{n=1}^\infty \sigma_{k-1}(n)  e^{-\frac{2\pi n i }{\tau}} \right]\\
&\notag =-\frac{2}{(k-1)!} S^{\text{P}}_{0,1-k}(q) + \tau^{-k}\mathcal{E}_0\left(k,0^{-1}; -\frac{1}{\tau}\right)\,.
\end{align}
It is simple to realise that 
\begin{equation}
S^{\text{P}}_{0,1-k}(q) = (q\partial_q)^{k-1} S^{\text{P}}_{k-1,0}(q)  = (2\pi i)^{1-k} \partial_\tau^{k-1} S^{\text{P}}_{k-1,0}(\tau)=\frac{\zeta(k)\Gamma(k)}{(2\pi i \tau)^{k}} -\frac{\zeta(2-k)}{2\pi i \tau}-  \frac{\zeta(k)\Gamma(k)}{(2\pi i)^k}
\end{equation}
where we notice that the second term is non-vanishing only for $k=2$ (remember $k$ is an even positive integer).
Rearranging the terms we obtain
\begin{equation}
G^0_k(\tau) + 2 \zeta(k) = \tau^{-k} \Big[G^0_k\Big(-\frac{1}{\tau}\Big)+2\zeta(k)\Big] + \frac{2 \pi i }{\tau} \delta_{k,2}\,,
\end{equation}
which is precisely telling us that the holomorphic Eisenstein series $G_k(\tau) =G^0_k(\tau) + 2 \zeta(k)$ are modular forms of weight $k$ for $k\geq 4$ even and quasi-modular for $k=2$, as we had already derived before in section \ref{sec:Eis}.

To understand the modular properties of the iterated integrals $\mathcal{E}_0(k,0^{k-2-\ell}; \tau)$ we now have three interesting intervals to consider: $\ell= -\beta \in \{0,...,k-2\}$, $\ell=-\beta \geq k-1$ and finally $\ell =-\beta<0$.

The first two intervals are the easiest to understand and the key property is that when the parameter $\beta$ is a negative integer the hypergeometric function appearing in (\ref{eq:SabTS}) is actually a polynomial of degree $\ell$ for $\ell=-\beta\in \{0,...,k-2\}$ or of degree $\ell -(k-1)$  for $\ell \geq k-1$.
In particular, given the discussion above, we notice that for $\ell \geq k-1$ we are actually just taking $l-(k-1)$ derivatives of the holomorphic Eisenstein series $G_k(-1/\tau)$ which is a modular function, hence the modularity properties will be ``spoilt'' by the derivative but they are very simple to recover.
 
 For the case $\mathcal{E}_0(k,0^{k-2-\ell}; \tau)$ with $\ell=-\beta \in \{0,...,k-2\}$ we can just expand the hypergeometric function in (\ref{eq:SabTS}) to obtain
 \begin{align}
 \label{eq:Sdual}
 \mathcal{E}_0(k,0^{k-2-\ell}; \tau) &= \sum_{p=0}^\ell \frac{\Gamma(k-1-p)}{\Gamma(k-1-\ell)} \binom{\ell}{p} \frac{\tau^{k-2-\ell-p}}{(2\pi i)^{\ell-p}}  \mathcal{E}_0\left(k,0^{k-2-p}; -\frac{1}{\tau}\right)\\
 &\quad\notag -\frac{2}{\Gamma(k) (2\pi i)^\ell} \partial_\tau^{\ell}  S^{\text{P}}_{k-1,0}(q)\,,
 \end{align}
 and note that $(2\pi i)^{-\ell}  \partial_\tau^{\ell}  S^{\text{P}}_{k-1,0}(q) = S^{\text{P}}_{k-1-\ell,\ell}(q)$.
We can also invert the above expression
 \begin{align}
 \mathcal{E}_0\left(k,0^{k-2-\ell}; -\frac{1}{\tau}\right) &= \sum_{p=0}^\ell (-1)^{\ell+p} \frac{\Gamma(k-1-p)}{\Gamma(k-1-\ell)} \binom{\ell}{p} \frac{\tau^{\ell+p+2-k}}{(2\pi i)^{\ell}} \bigg[  (2\pi i)^p \mathcal{E}_0(k,0^{k-2-p}; \tau)\\
 &\quad\quad\quad +  \frac{2}{\Gamma(k)} \partial_\tau^{p}  S^{\text{P}}_{k-1,0}(q)\bigg]\,,\nn
 \end{align}
which reduces to the two special cases discussed above for $\ell=0$ and $\ell=k-1$.

This equation nicely exhibits an upper triangular structure where the $S$-transformation of $\mathcal{E}_0(k,0^{k-2-\ell};\tau)$ involves all other iterated integrals $\mathcal{E}_0(k,0^{k-2-p};\tau)$ with $p\in \{0,...,\ell\}$. It should thus be thought of as a finite upper triangular matrix. If we consider all the $\mathcal{E}_0(k,0^{k-2-\ell};\tau)$ with $\ell \in \{0,...,k-2\}$ as a $(k-1)$ dimensional vector, we then have that under $S$-transformation this vector transforms with a $(k-1)\times (k-1)$ upper triangular matrix (plus a $(k-1)$ vector of cocycles, i.e. the Laurent polynomials in $\tau$). This vector can be thought of as arising from a $(k-1)$ dimensional representation of $SL_2(\mathbb{R})$ \cite{Brown:2014,Brown:2017}.

This finiteness property of the S duality transformation does not happen in the remaining case to analyse, i.e. the case $\ell = -\beta <0$, for which it is worth noticing that this range of parameters lies outside the class analysed in \cite{Brown:2014}.
Let us consider for example $\mathcal{E}_0(4,0,0,0; \tau)$ for which $\alpha=4$ and $\beta = 1$. Using (\ref{eq:SabTS}) we have
 \begin{align}
 \mathcal{E}_0(4,0,0,0; \tau) &= \mathcal{E}_0^{\text{P}}(4,0,0,0; \tau)-\frac{2}{3!}\tau^{4} \sum_{n=1}^\infty \frac{\sigma_{-3}(n)}{n}  e^{-\frac{2\pi n i }{\tau}}\,_2F_0\left(4,1;\frac{i\tau}{2\pi n}\right)\nn\\
 &= \mathcal{E}_0^{\text{P}}(4,0,0,0; \tau) + \tau^4\sum_{p=0}^\infty (4)_p \left(\frac{i\tau}{2\pi}\right)^p\mathcal{E}_0\left(4,0^{3+p}; -\frac{1}{\tau}\right)\,,
 \end{align}
where the perturbative part $\mathcal{E}_0^{\text{P}}(4,0,0,0; \tau)$ was presented in (\ref{eq:Ex400}).
It is manifest that the $S$-transform of an iterated integral outside the range $\ell\leq k-2$ produces an infinite tower of higher and higher iterated integrals, it is only for $ \mathcal{E}_0(k,0^{k-2-\ell}; \tau)$ with $\ell\geq 0$ that we produce a finite upper triangular tower.  As we have already mentioned this case does not fall within the class of objects studied in \cite{Brown:2014,Brown:2017}, however it is tantalising to think that perhaps this case might correspond to an $\infty$-dimensional representation of $SL_2(\mathbb{R})$ (such as a Verma module) rather than a finite-dimensional one.

We notice that our dictionary (\ref{eq:EklS}) between iterated integrals and the q-series studied is still meaningful even when we consider $k\in\mathbb{N}$ odd, for example by slight abuse of notation $ \mathcal{E}_0(3,0;\tau)$ can still be understood as genuine iterated integral of the q-series $\Lam_{-2}(q) = \sum_{n\geq1} \sigma_{2}(n) q^n$ as in  (\ref{Lambert2}).

We can then consider $ \mathcal{E}_0(k,0^{k-2};\tau)$ with $k\in\mathbb{N}$ odd and proceed as above to derive its modular property under $\tau\to \pm S\tau$. The only difference with the $k$ even case (\ref{eq:E0Lagom}) lies in the fact that the perturbative series does not truncate in this case.
Using (\ref{eq:QuasiMod})-(\ref{eq:QuasiMod2}) we know that the new ``cocycles'' will not be Laurent polynomials (\ref{eq:cocycle}) any longer but rather they will be given by the multi-valued lateral resummations (\ref{eq:SabFullGen}) $\mathcal{S}_{\mp}\left[ S^T_{k-1,0}\right](\tau)$. It would be extremely interesting to understand whether one can generalise the cohomological arguments of \cite{ZagierFR} to accommodate for this more general case of multi-valued ``cocycles''.

As a last comment for this section, similar to the observation at the end of section~\ref{sec:Lambert}, we note that the $q$-series $S_{\alpha,\beta}(q)$ defined in~\eqref{eq:Sab} has weight given by $k=1-\alpha-\beta$ and it is an eigenfunction of all Hecke operators $T_n$ that were defined in~\eqref{eq:Heckedef}. Using the property~\eqref{eq:Heckenec} and the expansion~\eqref{eq:SabItInt} it is easy to show that
\begin{align}
T_n S_{\alpha,\beta} = n^{-\beta} \sigma_{\beta-\alpha}(n) S_{\alpha,\beta} = n^{-\alpha} \sigma_{\alpha-\beta}(n) S_{\alpha,\beta}\,,
\end{align}
for any $\alpha$ and $\beta$, and clearly $S_{\alpha,\beta}$ is Hecke normalised. Given the relation to iterated Eisenstein integrals of depth one stated in~\eqref{eq:EklS}, this means that we have also
\begin{align}
T_n \mathcal{E}_0(k, 0^{k-2-\ell};\tau) = n^{\ell+1-k} \sigma_{k-1}(n) \mathcal{E}_0(k, 0^{k-2-\ell};\tau)\,.
\end{align}
Some connections between Hecke operators and iterated integrals were explored in \cite{Brown:2017b}.

\section{Single-valued prescription}
\label{sec:SV}

In this final section, we study the relation between our results and the elliptic analogue of the single-valued map (denoted $\esv$ in what follows)~\cite{Zerbini:2018sox,Broedel:2018izr,Gerken:2018jrq,Zagier:2019eus,Gerken:2019cxz}. From a string theory perspective, the map $\esv$ is meant to provide the one-loop generalisation of the single-valued relation between open and closed string tree-level amplitudes~\cite{Schlotterer:2012ny,Schnetz:2013hqa,Brown:2013gia,Stieberger:2013wea,Stieberger:2014hba,Schlotterer:2018zce,Brown:2018omk,Brown:2019wna}. Unlike the tree-level case the exact form of the map $\esv$ is unknown with conjectured pieces of it given in~\cite{Broedel:2018izr,Gerken:2018jrq,Gerken:2019cxz}, and for the present discussion we will mainly use the ingredients of~\cite{Broedel:2018izr,Gerken:2018jrq}. The $q$-series $S_{\alpha,\beta}(q)$ was related to holomorphic Eisenstein series for special values of $\alpha$ and $\beta$, but also produced generalisations thereof. We shall now investigate how the images of $S_{\alpha,\beta}$ under the elliptic single-valued map are related to non-holomorphic Eisenstein series.

To explain expressions to which the map $\esv$ can be applied, we begin, following~\cite{Broedel:2018izr}, with an iterated integral over the A-cycle of the genus-one string torus, apply an $S$-duality transformation to obtain an iterated integral over the B-cycle. Here it is important to start with zero weight expressions which at depth one means according to (\ref{eq:SabSDual}) that we need to consider the case $\alpha+\beta= 1$, or equivalently $\alpha=k,\,\beta=1-k$ and study
\begin{equation}
 \mathcal{E}_0(2k,0^{k-1 }; \tau) = -\frac{2}{\Gamma(2k)} S_{k,1-k}(\tau)\,,
\end{equation}
where for the moment we will consider $k\in\mathbb{N}\setminus\{0\}$. The above can be thought of as the `balanced, weight zero, middle case' in terms of possible iterated integrals.
Let us first consider
\begin{equation}
-\frac{\Gamma(2k)}{\Gamma(k)} \mathcal{E}_0\left( 2k,0^{k-1 }; -\frac{1}{\tau}\right) = \frac{2}{\Gamma(k)} S_{k,1-k}\left(-\frac{1}{\tau}\right)\,,
\end{equation}
using the general formula (\ref{eq:Sdual}) and then apply the $\esv$ prescription.
Specialising (\ref{eq:Sdual}) to $k\to 2k$ and $\ell\to k-1$ we have
 \begin{align}
-\frac{\Gamma(2k)}{\Gamma(k)} \mathcal{E}_0\left( 2k,0^{k-1 }; -\frac{1}{\tau}\right) &=-\frac{\Gamma(2k)}{\Gamma(k)} \sum_{p=0}^{k-1} \frac{\Gamma(2k-1-p)}{\Gamma(k)} \binom{k-1}{p}(-2\pi i \tau)^{p+1-k} \mathcal{E}_0(2k,0^{2k-2-p}; \tau)\nn\\
 &\quad +\frac{2}{\Gamma(k) }  S^{\text{P}}_{k,1-k}\left(-\frac{1}{\tau}\right)\,,
 \end{align}
 wich can be rewritten as 
 \begin{align}\label{eq:SVTot}
-\frac{\Gamma(2k)}{\Gamma(k)} \mathcal{E}_0\left(2k,0^{k-1 }; -\frac{1}{\tau}\right) =&-\Gamma(2k) (-2\pi i \tau) \sum_{p=0}^{k-1} \binom{2k-2-p}{k-1 }\frac{ (-2\pi i \tau)^{p-k}}{p!} \mathcal{E}_0(2k,0^{2k-2-p}; \tau)\nn\\
 & +\frac{2}{\Gamma(k) }  S^{\text{P}}_{k,1-k}\left(-\frac{1}{\tau}\right)\,.
 \end{align}
 
Before applying the $\esv$ prescription let us focus on the purely perturbative truncating part obtained from (\ref{eq:AsySab}):
\begin{align}
S^{\text{P}}_{k,1-k}\left(-\frac{1}{\tau}\right) &=  \left(\frac{2\pi i }{\tau}\right)^{k-1}\Gamma(1-k)\zeta(2-2k)\nn\\
&\quad\quad +\left(\frac{\tau}{2\pi i}\right)^k \Gamma(k)\zeta(2k)+ \sum_{n=0}^k  \frac{(-2\pi i/ \tau)^n}{n!} \zeta(k-n)\zeta(1-k-n)\,.
\end{align}
At the present time we are interested in the case of $k\in\mathbb{N}$ hence the first term and the $n=k-1$ in the sum have to be rewritten making use of Riemann's functional equation to arrive at an expression containing only zetas at positive arguments given by
 \begin{align}
\label{eq:SPsv}
&\quad\quad \frac{2}{\Gamma(k)} S^{\text{P}}_{k,1-k}\left(-\frac{1}{\tau}\right) \\
&\notag=\frac{(2k-3)!\,4^{2-k}}{(k-2)!(k-1)!} \left(\frac{-i \pi \tau}{2}\right)^{1-k} \zeta(2k-1)(1+\zeta(0)) +(-1)^{k-1} \frac{B_{2k}}{(2k)!} 4^k \left(\frac{-i \pi \tau}{2}\right)^k \\
&\notag \phantom{=}+2^{2-k} \sum_{n=0}^{k-2} \frac{(k)_n }{n!} \cos\left(  \frac{\pi(k+n)}{2}\right)\frac{\zeta(k-n)\zeta(k+n)}{\pi^{k-n}\, (i \pi \tau)^{n}} -2^{1-k} \frac{(2k-1)!}{k!\, (k-1)!} \frac{ \zeta(2k)}{(- i \pi \tau)^{k}}\,.
\end{align}
The $\zeta(0)$ part of the first term comes precisely from the $n=k-1$ term in the perturbative sum, while the last term in the above expression comes from the $n=k$ term.

The (conjectural) $\esv$ prescription amounts to (for $n>0$)~\cite{Broedel:2018izr,Gerken:2018jrq}\footnote{The variable $\yO$ agrees with $y$ in~\cite{Broedel:2018izr} and is introduced here to ease the comparison.}
 \begin{align}
 \label{eq:esv}
\esv:\quad \left\{\begin{array}{rl} \tau &\to\quad \tau-\bar{\tau} = 2 i \tau_2 = \frac{2 i \yO}{\pi}\,,\\
 \mathcal{E}_0 &\to\quad 2 \mbox{Re}\,\mathcal{E}_0\,,\\
\zeta(2n) &\to\quad 0\,,\\
\zeta(2n+1) &\to\quad 2 \zeta(2n+1)\,.
\end{array}\right.
 \end{align}

Note that in all above expressions we used the combination $\frac{-i \pi \tau}{2}$ so that $\esv\left(\frac{-i \pi \tau}{2}\right) = \yO$; hence we see that once we apply $\esv$ to (\ref{eq:SPsv}) only the first two terms survive: in the sum the cosine forces $k$ and $n$ to have the same parity, which implies that both $\zeta(k\pm n)$ are even zetas and the last term is directly an even zeta.
Putting (\ref{eq:SVTot}) together with (\ref{eq:SPsv}) we arrive at
\begin{align}
\label{eq:esvIntk}
&\quad  \esv\Big[ -\frac{\Gamma(2k)}{\Gamma(k)} \mathcal{E}_0\left(2k,0^{k-1 }; -\frac{1}{\tau}\right) \Big]\nn\\
&\notag= \frac{4(2k-3)!}{(k-2)!(k-1)!} \zeta(2k-1) \left(4 \yO \right)^{1-k} +(-1)^{k-1} \frac{B_{2k}}{(2k)!}  \left(4 \yO\right)^k\\
&\quad\notag\phantom{=}- 8 \yO \Gamma(2k) \sum_{p=0}^{k-1} \binom{2k-2-p}{k-1}\frac{ (4\yO)^{p-k}}{p!} \mbox{Re}\left[ \mathcal{E}_0(2k,0^{2k-2-p}; \tau)\right] \\
&=E_k(\tau) \,,
\end{align}
where $E_k$ denotes now the non-holomorphic Eisenstein series in the normalisation
\begin{align}
E_k(\tau) &= \pi^{-k} \sum_{(c,d)\neq (0,0)} \frac{(\Im\,\tau)^k}{|c\tau+d|^{2k}}\nn\\
&= \frac{2\zeta(2k)}{\pi^k}  \tau_2^k + \frac{2 \Gamma(k-\frac{1}{2}) \zeta(2k-1)}{\Gamma(k) \pi^{k-1/2}} \tau_2^{1-k} \nn\\
&\hspace{30mm}+\frac{4  \sqrt{\tau_2}}{\Gamma(k)}\sum_{n\in\mathbb{Z}\setminus\{0\}}\sigma_{1-2k}(\vert n\vert )  \,\vert n \vert^{k-\frac{1}{2}} e^{ 2\pi i n \tau_1}  K_{k-\frac{1}{2}} ( 2\pi |n| \tau_2)\,.
\end{align}
In the above formula we have shown the general Fourier expansion of the non-holomorphic Eisenstein series that is valid for any value of $k$. If $k$ is an integer the Bessel function has an exact expansion for large $\tau_2$ that gives a rise to an explicit series in $q^n+\bar{q}^n$ that agrees with the one of the real part of the iterated Eisenstein integral. Our expression~\eqref{eq:esvIntk} matches precisely equation (2.34) of~\cite{Broedel:2018izr}. 

Let us try and repeat this analysis for the case $k$ generic. 
In this case the expression $\mathcal{E}_0(2k,0^{k-1 }; -\frac{1}{\tau})$ cannot really be thought anymore of a proper iterated integral, e.g. for $k=3/2$ we would have $\mathcal{E}_0(3,0^{\frac{1}{2}}; -\frac{1}{\tau})$ so a fractional number of integrations. We will however interpret this as its associated $q$-series expansion and start again from the expression in terms of $S_{k,1-k}$ via
\begin{equation}
-\frac{\Gamma(2k)}{\Gamma(k)} \mathcal{E}_0\left( 2k,0^{k-1 }; -\frac{1}{\tau}\right) = \frac{2}{\Gamma(k)} S_{k,1-k}\left(-\frac{1}{\tau}\right)\,.
\end{equation}
We have to consider now the general case (\ref{eq:SabTSGen}) specialised to $\alpha = k,\,\beta=1-k$.
Let us begin with the non-perturbative terms
\begin{align}
\frac{2}{\Gamma(k)} S_{k,1-k}^{\text{NP}}\left(-\frac{1}{\tau}\right) &\label{eq:SNPU}= \frac{2}{\Gamma(k)} \sum_{n=1}^\infty \frac{\sigma_{1-2k}(n)}{n^{1-k}}  e^{2\pi n i \tau }\,_2F_0\left(k,1-k;-\frac{i}{2\pi \tau n}\right)\\
&
\notag=\frac{2}{\Gamma(k)} \sum_{n=1}^\infty \frac{\sigma_{1-2k}(n)}{n^{1-k} }  q^n \left(-2\pi i n \tau \right)^{1-k} \,U(1-k,2-2k; -2\pi i n \tau)\,.
\end{align}
We can apply the $\esv$ prescription is a simple way now via $\tau \to 2 i \tau_2$ and $q^n\to q^n +\bar{q}^n$ (note that for this prescription we think of $q$ as independent of $\tau$) and after rearranging the sum we obtain 
\begin{align}
& \quad \esv\left[ \frac{2}{\Gamma(k)} S_{k,1-k}^{\text{NP}}\left(-\frac{1}{\tau}\right) \right] \nn\\
&= \frac{2}{\Gamma(k)} \sum_{ n\in\mathbb{Z}{\setminus\{0\}} } \frac{\sigma_{1-2k}(\vert n\vert )}{|n|^{1-k}} e^{ 2\pi i n \tau_1} e^{ -2\pi   |n| \tau_2}  \left(4 \pi |n| \tau_2 \right)^{1-k} \,U(1-k,2-2k; 4 \pi |n| \tau_2 )\nn\\
&= \frac{4  \sqrt{\tau_2}}{\Gamma(k)}\sum_{n\in \mathbb{Z} {\setminus\{0\}} }\sigma_{1-2k}(\vert n\vert )  \,\vert n \vert^{k-\frac{1}{2}} e^{ 2\pi i n \tau_1}  K_{k-\frac{1}{2}} ( 2\pi |n| \tau_2) \,,
\end{align}
where we rewrote the confluent hypergometric function $U$ in terms of modified Bessel function of the second kind $K$. For $k$ integer the Bessel function reduces to a spherical Bessel and can actually be written as an exponentially suppressed term $e^{-2\pi n \tau_2}$ times a polynomial of degree $k-1$ in $1/\tau_2$. This is the reason why in the single-valued prescription for the $k$ integer case, given by equation (\ref{eq:esvIntk}), we only have a finite sum over iterated integrals.

For the present case of $k$ generic the Bessel function does not truncate but it is still exponentially suppressed as $\tau_2\to +\infty$.
Let us move to the first two perturbative terms in (\ref{eq:SabTSGen}) which after the $S$ duality transformation $\tau \to -1/\tau$ are of the form
\begin{align}
&\quad\quad \frac{2}{\Gamma(k)}  \left(\frac{2\pi i }{\tau}\right)^{k-1}\Gamma(1-k)\zeta(2-2k)+\frac{2}{\Gamma(k)}  \left(\frac{\tau}{2\pi i}\right)^k \Gamma(k)\zeta(2k) \\
&=\notag \frac{2 \Gamma(k-\frac{1}{2}) \zeta(2k-1)}{\Gamma(k) \pi^{k-1/2}} \left(\frac{-i\tau}{2}\right)^{1-k} + \frac{2\zeta(2k)}{\pi^k}  \left(\frac{-i\tau}{2}\right)^k\,,
\end{align}
similarly to what we obtained above for the integer $k$ case but with two major differences. Firstly we notice that for $k$ generic $\zeta(2k-1)$ and $\zeta(2k)$ will remain untouched, and we are not aware of a similar extension of the single-valued map applied to the zeta function at a generic argument. Secondly we see that the term containing $\zeta(0)$ in (\ref{eq:SPsv}) is now absent from the asymptotic perturbative series (\ref{eq:SabFullGen})
\begin{align}
\mathcal{S}_-[S^T_{k,1-k}]\left(-\frac{1}{\tau}\right) &= \sum_{n=1}^\infty \frac{\sigma_{1-2k}(n)}{n^{1-k} }   \left(-2\pi i n \tau \right)^{1-k} \,U(1-k,2-2k; -2\pi i n \tau) \\
&\notag \sim \sum_{m=0}^\infty \frac{(-2\pi i / \tau)^m }{m!}\zeta(k-m)\zeta(1-k-m)\,,
\end{align}
where we evaluated the Borel resummation (\ref{eq:SabFullGen}) for the particular case $\alpha=k\,,\beta=1-k$, which we know is asymptotic to (\ref{eq:AsySab}).
We note how similar the resummed asymptotic perturbative series is to the non-perturbative terms (\ref{eq:SNPU}) with the only difference given by the absence of the $q^n$ term in the summand.

In this general $k$ case we are not aware of any argument why the $\esv$ prescription should remove the resummed perturbative expansion, hence we obtain the expression 
\begin{align}
&\quad\,\, \notag \esv \Big[ \frac{2}{\Gamma(k)} S_{k,1-k}\left(-\frac{1}{\tau}\right)\Big]\\
&\notag= \esv \Big[\frac{2 \Gamma(k-\frac{1}{2}) \zeta(2k-1)}{\Gamma(k) \pi^{k-1/2}} \left(\frac{-i\tau}{2}\right)^{1-k} + \frac{2\zeta(2k)}{\pi^k}  \left(\frac{-i\tau}{2}\right)^k +\frac{2}{\Gamma(k)} S_{k,1-k}^{\text{NP}}\left(-\frac{1}{\tau}\right) \\
&\notag \phantom{=\esv\,\,\,}+  \frac{2}{\Gamma(k)} \mathcal{S}_-[S^T_{k,1-k}]\left(-\frac{1}{\tau}\right) \Big]\nn\\
&\notag= \frac{2 \Gamma(k-\frac{1}{2}) \zeta(2k-1)}{\Gamma(k) \pi^{k-1/2}} \tau_2^{1-k} + \frac{2\zeta(2k)}{\pi^k}  \tau_2^k + \frac{4  \sqrt{\tau_2}}{\Gamma(k)}\sum_{n\in\mathbb{Z}\setminus\{0\}}\sigma_{1-2k}(\vert n\vert )  \,\vert n \vert^{k-\frac{1}{2}} e^{ 2\pi i n \tau_1}  K_{k-\frac{1}{2}} ( 2\pi |n| \tau_2)\\
&\notag\phantom{=\esv\,\,\,}+\esv\Big[  \frac{2}{\Gamma(k)} \mathcal{S}_-[S^T_{k,1-k}]\left(-\frac{1}{\tau}\right)\Big] \\
&= E_k(\tau)+ \esv\Big[  \frac{2}{\Gamma(k)} \mathcal{S}_-[S^T_{k,1-k}]\left(-\frac{1}{\tau}\right)\Big]\,,
\end{align}
where again $E_k(\tau)$ denotes the non-holomorphic Eisenstein series for any $k$.
It is interesting to notice that the case $k$ half-integer has to be understood as a limiting case of the generic case, in particular the coefficients of the term $\tau_2^{1-k}$ despite being now a zeta even $ \zeta(2k-1)$ is not projected to zero, while the coefficient of the $\tau_2^k$ term despite being now a zeta odd $\zeta(2k)$ is not doubled by the esv prescription.

The case $k$ integer is very subtle since both the $\tau_2^{1-k}$ and $\tau_2^k$ have non-trivial $\esv$ projections  and must be combined with $\esv \,\mathcal{S}_-[S^T_{k,1-k}]$ to produce equation (\ref{eq:SPsv}).
It seems that for generic $k$ it is only the multi-valued analytic function $\mathcal{S}_-[S^T_{k,1-k}]$ that produces an obstruction to the equivariance under $SL_2(\ints)$ of $\esv\,S_{k,1-k}$ in the sense of \cite{Brown:2017}.
It would be extremely interesting to understand the role played by the resummed perturbative series $\mathcal{S}_-[S^T_{k,1-k}]$ from an iterated integral point of view and possibly find a modified esv prescription that would take care of it.

Finally we notice how the functional equation $\Gamma(k)E_k(\tau) = \Gamma(1-k)E_{1-k}(\tau)$ can be easily derived using the symmetry of $S_{k,1-k}(\tau) = S_{1-k,k}(\tau)$.

 \appendix
 
 \section{\texorpdfstring{An application to $q$-Pochhammers}{An application to q-Pochhammers}}
 
In this appendix, as a further check of our proposed transseries expansion, we analyse a particular case of the Lambert series (\ref{Lambert}) directly related to a certain $q$-Pochhammer symbol.

The $q$-Pochhammer symbol is defined as the infinite product
\begin{equation}\label{q-Pochhammer}
(a,q)_\infty = \prod_{n=0}^\infty (1-a q^n).
\end{equation}
By setting $a=q$, the $q$-Pochhammer symbol can be related to the Lambert series at $s=1$.
The proof is straightforward once we use $\Li_1(q) = - \log(1-q)$
\begin{align}
\Lam_1(q)  &= \sum_{n=1}^\infty \text{Li}_1(q^{n}) = - \sum_{n=1}^\infty \log(1-q^n) =-\log \prod_{m=0}^\infty \left(1-q\cdot q^m\right) = -\log (q,q)_\infty \,,
\end{align}
hence we have
\begin{equation}
(q,q)_\infty = e^{ - \Lam_1(q)}\,.\label{eq:Poch}
\end{equation}

The $q$-Pochhammer symbol is also closely related to the Dedekind $\eta$ function via
\begin{equation}
(q,q)_\infty = q^{-1/24} \eta(\tau)\,,\label{eq:Qeta}
\end{equation}
where $\tau$ is the half-period ratio and $q =e^{2\pi i \tau}$ is the square of the nome. 

Since equation (\ref{eq:Poch}) relates the $q$-Pochhammer $(q,q)_\infty$ to the Lambert series $\Lam_1(q)$, we need to specialise equation (\ref{eq:TSOdd}) to the $m=1$ case and obtain 
\begin{equation}
\Lam_1(e^{-2\pi y} ) =\frac{\log y}{12} +\frac{\pi}{12} (y^{-1}-y) + \Lam_1(e^{-\frac{2\pi}{y} })\,,
\end{equation}
where again $q=\exp(-2\pi y)$.

Passing to $q$-Pochhammers we have
\begin{equation}
(e^{-2\pi y},e^{-2\pi y})_\infty =\frac{1}{\sqrt{y}} \, e^{\frac{\pi}{12} \left( y-y^{-1}\right) } \left(e^{-\frac{2\pi}{y}},\,e^{-\frac{2\pi}{y}}\right)_\infty\,,\label{eq:PochQy}
\end{equation}
or equivalently going back to the $q$ variable
\begin{equation}\label{eq:PochQ}
(q,q)_\infty = \sqrt{\frac{2\pi}{\log(1/q)}} \, e^{\frac{\pi^2}{6 \log q}} \,q^{-\frac{1}{24}} \left(e^{\frac{4\pi^2}{\log q}},\,e^{\frac{4\pi^2}{\log q}}\right)_\infty\,.
\end{equation}
Note that for $q\to 1^-$ we have $e^{\frac{4\pi^2}{\log q}} \to 0$ and since $(0,0)_\infty = 1$ we have that (\ref{eq:PochQ}) can be approximated by
\begin{equation}\label{eq:PochQ2}
(q,q)_\infty \simeq \sqrt{\frac{2\pi}{\log(1/q)}} e^{\frac{\pi^2}{6 \log q}} q^{-\frac{1}{24}} \,,
\end{equation}
as already derived in \cite{Banerjee}.

Using formula (\ref{eq:Qeta}) we see that the proposed transseries does indeed produce the correct transformation under the action of the modular group.
Starting from (\ref{eq:PochQy}) we use $y= -i \tau$ so that
\begin{equation}
(e^{2\pi i \tau},e^{2\pi i \tau})_\infty =\frac{1}{\sqrt{-i \tau }} \, e^{\frac{ i \pi}{12} \left( \tau+\tau^{-1}\right) } \left(e^{-\frac{2\pi i}{\tau}},\,e^{-\frac{2\pi i}{\tau}}\right)_\infty\,,\label{eq:PochQtau}
\end{equation}
which we can rewrite in terms of $\eta(\tau)$ from (\ref{eq:Qeta})
\begin{equation}
q^{-\frac{1}{24}}\, \eta(\tau) = \frac{1}{\sqrt{-i \tau }}\, q^{-\frac{1}{24}} \, \eta\left( -\frac{1}{\tau}\right)\,,
\end{equation}
trivially reproducing the known $S$-transformation for the Dedekind $\eta$:
\begin{equation}
\eta(\tau) = \frac{1}{\sqrt{-i \tau }} \, \eta\left(-\frac{1}{\tau}\right)\,.
\end{equation}
This supports our claim that the transseries parameter does indeed exponentiate the way we presented.

\section{Dirichlet characters and Hurwitz sums}
\label{app:HurDir}
 
In this appendix, we derive some of the intermediate identities used in section~\ref{sec:otherroots} when studying the limit of the Lambert series $\Lam_s(q)$ for $q$ approaching a rational root of unity $e^{-2\pi y+2\pi i p/c}$ with $y\to 0^+$ and $p$ and $c$ co-prime integers. 

Our first object of study appears in~\eqref{eq:Hurconv} that we recall here for convenience:
\begin{align}
\sum_{\bar{h}=1}^{c-1}\sum_{l=1}^{c-1}  \zeta\left(k,\frac{l}{c}\right)\zeta\left(k-s,\frac{\bar{h}}{c}\right)  e^{ \pm 2\pi i \frac{\bar{h} l p^{-1} }{c} }\,.
\end{align}
The Hurwitz zeta has the following Dirichlet series representation valid for $(c,l)=1$ 
 \begin{equation}
 \label{eq:HurasDir}
 \zeta\left(k,\frac{l}{c}\right) = \frac{c^k}{\phi(c)}\sum_{\chi} \bar{\chi}(l) L(\chi,k)\,,
 \end{equation}
 where $\phi(c)$ denotes the Euler totient function of $c$ and the sum runs over all the Dirichlet characters modulo $c$, denoted by $\chi$, while $L(\chi,k)$ is the associated $L$-series, i.e. $L(\chi,k) = \sum_{n=1}^\infty \chi(n) n^{-k}$.
 
To avoid dealing with non-trivial subgroups of the cyclic group modulo $c$ we will assume that $c$ is a prime number, hence the number of Dirichlet characters modulo $c$ is precisely $\phi(c)=c-1$ since every $l \in \{1,...,c-1\}$ is coprime with $c$, i.e. $(c,l)=1$.
Making use of~\eqref{eq:HurasDir} we can then write 
\begin{align}
&\sum_{\bar{h}=1}^{c-1}\sum_{l=1}^{c-1}  \zeta\!\left(k,\frac{l}{c}\right)\zeta\!\left(k-s,\frac{\bar{h}}{c}\right)  e^{ \pm 2\pi i \frac{\bar{h} l p^{-1} }{c} }= \sum_{\chi,\chi'} \frac{c^{2k-s}}{\phi(c)^2} L(\chi,k)L(\chi',k-s)\sum_{\bar{h},l=1}^{c-1} \bar{\chi}(l)e^{ \pm 2\pi i \frac{\bar{h} l p^{-1} }{c} } \bar{\chi}'(\bar{h}).
\end{align}
First we notice that the sum over $\bar{h},l$ imposes $\chi = \chi'$ hence the sum over $(\chi,\chi')$ reduces to a single sum over all the Dirichlet characters $\chi_i$ with $i\in\{0,...,\phi(c)-1\}$. 
To prove this orthogonality of Dirichlet characters in Fourier space we take an integer $j$ such that $(j,c)=1$ and consider
\begin{align}
&\bar{\chi}(j) \sum_{\bar{h},l=1}^{c-1} \bar{\chi}(l)e^{ \pm 2\pi i \frac{\bar{h} l p^{-1} }{c} } \bar{\chi}'(\bar{h})  =\sum_{\bar{h},l=1}^{c-1} \bar{\chi}( j l)e^{ \pm 2\pi i \frac{\bar{h} l p^{-1} }{c} } \bar{\chi}'(\bar{h})\\
&\notag = \sum_{\bar{h},l'=1}^{c-1} \bar{\chi}( l')e^{ \pm 2\pi i \frac{\bar{h} l' j^{-1} p^{-1} }{c} } \bar{\chi}'(\bar{h})= \sum_{\bar{h}',l'=1}^{c-1} \bar{\chi}( l')e^{ \pm 2\pi i \frac{\bar{h}' l' p^{-1} }{c} } \bar{\chi}'(\bar{h}' j)\\
&=\notag \bar{\chi}'(j) \sum_{\bar{h}',l'=1}^{c-1} \bar{\chi}( l')e^{ \pm 2\pi i \frac{\bar{h}' l' p^{-1} }{c} } \bar{\chi}'(\bar{h}' )\,,
\end{align}
so that we must have $\chi = \chi'$.
In the above argument first we made use of the complete multiplicative properties of the Dirichlet characters and then changed summation variables to $l' \equiv l j (\mod c)$ and $\bar{h}' \equiv \bar{h} j^{-1} (\mod c)$, with $j^{-1}$ the multiplicative inverse of $j$ modulo $c$.

At this point,  we can use the completely multiplicative property of the Dirichlet characters for the associated L-series
\begin{equation}
L(\chi_i,k)L(\chi_i ,k-s) = \sum_{n,m\geq1} \frac{\chi_i(n)\chi_i(m)}{n^k m^{k-s}} = \sum_{N\geq1} \frac{\sigma_{-s}(N) \chi_i(N)}{N^{k-s}}\,.
\end{equation}

Putting everything together we obtain the expression
\begin{align}
\label{eq:Dirichletapp}
&\quad\quad \sum_{\bar{h}=1}^{c-1}\sum_{l=1}^{c-1}  \zeta\left(k,\frac{l}{c}\right)\zeta\left(k-s,\frac{\bar{h}}{c}\right)  e^{ \pm 2\pi i \frac{\bar{h} l p^{-1} }{c} }\nn\\
&= \sum_{N\geq1} \frac{\sigma_{-s}(N) c^{2k-s}}{N^{k-s}} \frac{1}{\phi(c)^{2}}\sum_{i=0}^{\phi(c)-1}\chi_i(N) \sum_{\bar{h},l=1}^{c-1} \bar{\chi_i}(l)e^{ \pm 2\pi i \frac{\bar{h} l p^{-1} }{c} } \bar{\chi_i}(\bar{h})\,.
\end{align}

The final double sum can be simplified by
\begin{align}
\label{eq:char}
 \frac{1}{\phi(c)^{2}}\sum_{i=0}^{\phi(c)-1}\chi_i(N) \sum_{\bar{h},l=1}^{c-1} \bar{\chi_i}(l)e^{ \pm 2\pi i \frac{\bar{h} l p^{-1} }{c} } \bar{\chi_i}(\bar{h}) &= \frac{1}{\phi(c)^{2}}\sum_{i=0}^{\phi(c)-1}\chi_i(N) \sum_{h,l=1}^{c-1} \bar{\chi_i}(l)e^{ \pm 2\pi i\frac{ hl  }{c} } \bar{\chi_i}(h p) \nn\\
&\ =\frac{1}{\phi(c)^{2}}\sum_{i=0}^{\phi(c)-1}\chi_i(N)  \bar{\chi_i}(p)\sum_{h,l=1}^{c-1} \bar{\chi_i}(l)e^{ \pm 2\pi i \frac{ hl  }{c} } \bar{\chi_i}(h)\nn\\
&  =\frac{1}{\phi(c)^{2}}\sum_{i=0}^{\phi(c)-1}\chi_i(N p^{-1}) \sum_{h,l=1}^{c-1} \bar{\chi_i}(l)e^{ \pm 2\pi i \frac{ hl  }{c} } \bar{\chi_i}(h)\nn\\
&= \frac{1}{\phi(c)}\sum_{i=0}^{\phi(c)-1}\chi_i(N p^{-1}) \sum_{\tilde{h}=1}^{c-1} \bar{\chi_i}( h)e^{ \pm 2\pi i \frac{\tilde{h}  }{c} } \\
&\notag=\sum_{\tilde{h}=1}^{c-1} e^{ \pm 2\pi i \frac{\tilde{h}  }{c} } \frac{1}{\phi(c)} \sum_{\chi} \chi( Np^{-1}) \bar{\chi}(\tilde{h})  \nn\\
&= e^{ \pm 2\pi i \frac{ p^{-1}N }{c} } \chi_0(N)\nn\\
&\equiv {\chi}^{\pm}(N) \,,\nn
\end{align}
where the character $\chi_0(N)$ is equal to $0$ when $N\equiv 0 \mod c$ and $1$ otherwise.
In here we simply changed summation variables defining $ h l \equiv \tilde{h} \mod c$ and used the fact that $\sum_\chi \chi(n) \bar{\chi}(a) $ is equal to $1$ if $n\equiv a \mod c$ and $0$ otherwise.

\medskip

The next expression we want to simplify occurs in~\eqref{eq:NPother} and reads
%\begin{equation}
%\sum_{N=1}^\infty \sigma_{-s}(N) \chi^-(N) e^{-2\pi \frac{N}{c^2y}} = \sum_{N \neq0\, mod\,c} \sigma_{-s}(N) e^{ \pm 2\pi i \frac{ p^{-1} N}{c} } e^{-2\pi \frac{N}{c^2y}}\,,
%\end{equation}
%where we used the character defined in (\ref{eq:char}).
\begin{align}
\label{eq:charsum}
&\quad \quad \sum_{N=1}^\infty \sigma_{-s}(N) \chi^-(N) e^{-2\pi \frac{N}{c^2y}}\nn\\
 &= \sum_{N \neq0\, \mod\,c} \sigma_{-s}(N) e^{ - 2\pi i \frac{ p^{-1} N}{c} } e^{-2\pi \frac{N}{c^2 y}}\nn\\
&=\Lam_s\left( \frac{1}{c^2y} + i \,\frac{p^{-1}}{c}\right) -\sum_{N\geq1} \sigma_{-s}(cN)  e^{-2\pi \frac{N}{cy}}\\
&\notag= \Lam_s\left(\frac{1}{ c^2 y} + i\, \frac{p^{-1}}{c}\right) - (1+c^{-s})  \sum_{N \neq0\, \mod\,c} \sigma_{-s}(N)  e^{-2\pi \frac{N}{cy}}-\sum_{N\geq1} \sigma_{-s}(c^2N)  e^{-2\pi \frac{N}{y}}\\
&\notag = \Lam_s\left( \frac{1}{ c^2 y} + i\, \frac{p^{-1}}{c}\right) -(1+c^{-s}) \Lam_s(\frac{1}{ cy}) +\sum_{N \geq 1} \left[(1+c^{-s}) \sigma_{-s}( c N) -\sigma_{-s}( c^2N) \right] ^{-2\pi \frac{N}{y}}\\
&= \notag \Lam_s\left( \frac{1}{c^2 y} + i \,\frac{p^{-1}}{c}\right) -(1+c^{-s})\, \Lam_s\left(\frac{1}{cy}\right) + c^{-s} \Lam_s\left(\frac{1}{y}\right)\,,
\end{align}
where first we have written out the character defined in (\ref{eq:char}) and then used the fact that for $(c,N)=1$ and $c$ prime we have $\sigma_{-s}( cN) = (1+c^{-s}) \sigma_{-s}(N)$. Moreover, the following identity
\begin{equation}\label{eq:sigmaid}
(1+c^{a s})\sigma_{s}(c^a N) -\sigma_{s}(c^{2a}N) = c^{a s} \sigma_{s}(N)\,,
\end{equation}
valid for $c$ prime not necessarily coprime with $N$ and for any integer $a\geq0$ was used.


\begin{thebibliography}{40}


\bibitem{ZagierApp}
  D.~B.~Zagier,
  ``The Mellin transform and other useful analytic techniques,''
    Appendix to E.~Zeidler, \textit{Quantum Field Theory I: Basics in Mathematics and Physics. A Bridge Between Mathematicians and Physicists} (Springer, Berlin-Heidelberg-New York, 2006) 305--323.
   
\bibitem{Ecalle:1981}
   J.~Ecalle,  ``Les Fonctions Resurgentes,'' {\hypersetup{urlcolor=darkred}\href{https://www.math.u-psud.fr/~ecalle/publi.html}{vol.~I - III.
\newblock Publ. Math. Orsay, 1981}\hypersetup{urlcolor=blue}}.

\bibitem{delabaere1999resurgent}
E.~Delabaere and F.~Pham, 
 ``Resurgent methods in semi-classical asymptotics,''  
 {\hypersetup{urlcolor=darkred}\href{http://www.numdam.org/item/AIHPA_1999__71_1_1_0}{Ann. Inst. H. Poincar\'e Phys. Th\'eor. {\bf 71} (1999) 1}\hypersetup{urlcolor=blue}}.
  
  \bibitem{Dunne:2016jsr}
  G.~V.~Dunne and M.~\"Unsal,
  ``Deconstructing zero: resurgence, supersymmetry and complex saddles,''
  \doi{JHEP {\bf 1612} (2016) 002}{doi:10.1007/JHEP12(2016)002}
  \eprintN{1609.05770}.
  
  
\bibitem{Shimomura}
  S.~Shimomura, 
  ``Modularity gap for Eisenstein series,''
  \doi{Proc. Japan Acad. {\bf 86} (2010) 29--84}{doi: 10.3792/pjaa.86.79}.


\bibitem{Knopp} 
K. Knopp, ``Grenzwerte von Reihen bei der Ann\"{a}herung an die Konvergenzgrenze,'' Inaugural Dissertation, Berlin (1907), p. 34.

\bibitem{Banerjee} 
S.~Banerjee and B.~ Wilkerson ``Lambert series and q-functions near $q=1$,'' \eprintNT{1602.01085}.


  
  \bibitem{Brown:2014}
  F.~Brown, ``Multiple Modular Values and the relative completion of the fundamental group of $\mathcal{M}_{1,1}$,''
  \eprintNT{1407.5167v4}.
  
\bibitem{Broedel:2015hia}
  J.~Broedel, N.~Matthes and O.~Schlotterer,
  ``Relations between elliptic multiple zeta values and a special derivation algebra,''
  \doi{J.\ Phys.\ A {\bf 49} (2016) no.15,  155203}{doi:10.1088/1751-8113/49/15/155203}
  \eprintN{1507.02254}.
  %%CITATION = doi:10.1088/1751-8113/49/15/155203;%%
  
  
  \bibitem{Brown:2017qwo}
  F.~Brown,
  ``A class of non-holomorphic modular forms I,''
  \eprintNT{1707.01230}.
  %%CITATION = ARXIV:1707.01230;%%
  %22 citations counted in INSPIRE as of 13 Jan 2020
  
  \bibitem{Brown:2017}
  F.~Brown, ``A class of non-holomorphic modular forms II : equivariant iterated Eisenstein integrals,''
  \eprintNT{1708.03354}.
  
    \bibitem{Brown:2017b}
  F.~Brown, ``A class of non-holomorphic modular forms III : real analytic cusp forms for $SL_2(\mathbb{Z})$,''
  \eprintNT{1710.07912}.

\bibitem{Broedel:2018izr}
  J.~Broedel, O.~Schlotterer and F.~Zerbini,
  ``From elliptic multiple zeta values to modular graph functions: open and closed strings at one loop,''
  \doi{ JHEP {\bf 1901} (2019) 155}{doi:10.1007/JHEP01(2019)155}
  \eprintN{1803.00527}.


\bibitem{Broedel:2018iwv}
  J.~Broedel, C.~Duhr, F.~Dulat, B.~Penante and L.~Tancredi,
  ``Elliptic symbol calculus: from elliptic polylogarithms to iterated integrals of Eisenstein series,''
  \doi{JHEP {\bf 1808} (2018) 014}{doi:10.1007/JHEP08(2018)014}
  \eprintN{1803.10256}.
  %%CITATION = doi:10.1007/JHEP08(2018)014;%%

\bibitem{Richter:2018hun}
  G.~Richter,
  ``Iterated Integrals and genus-one open-string amplitudes,''
  \doi{PhD thesis, Humboldt University Berlin (2018)}{doi:10.18452/19309}.

\bibitem{Broedel:2019vjc}
  J.~Broedel and O.~Schlotterer,
  ``One-Loop String Scattering Amplitudes as Iterated Eisenstein Integrals,''
  \doi{Proceedings of KMPB Conference: Elliptic Integrals, Elliptic Functions and Modular Forms in Quantum Field Theory}{doi:10.1007/978-3-030-04480-0_7}.
  %%CITATION = doi:10.1007/978-3-030-04480-0_7;%%


\bibitem{DHoker:2015gmr}
  E.~D'Hoker, M.~B.~Green and P.~Vanhove,
  ``On the modular structure of the genus-one Type II superstring low energy expansion,''
  \doi{JHEP {\bf 1508} (2015) 041}{doi:10.1007/JHEP08(2015)041}
  \eprintN{1502.06698}.
  %%CITATION = doi:10.1007/JHEP08(2015)041;%%
   
\bibitem{DHoker:2015wxz}
  E.~D'Hoker, M.~B.~Green, \"{O}.~G\"{u}rdogan and P.~Vanhove,
  ``Modular Graph Functions,''
  \doi{Commun.\ Num.\ Theor.\ Phys.\  {\bf 11} (2017) 165}{doi:10.4310/CNTP.2017.v11.n1.a4}
  \eprintN{1512.06779}.
  %%CITATION = doi:10.4310/CNTP.2017.v11.n1.a4;%%

\bibitem{DHoker:2016mwo}
  E.~D'Hoker and M.~B.~Green,
  ``Identities between Modular Graph Forms,''
  J.\ Number Theor.\  {\bf 189} (2018) 25
  \eprintN{1603.00839}.
  %%CITATION = ARXIV:1603.00839;%%

\bibitem{Mafra:2019ddf}
  C.~R.~Mafra and O.~Schlotterer,
  ``All-order alpha'-expansion of one-loop open-string integrals,''
  \eprintN{1908.09848}.
  %%CITATION = ARXIV:1908.09848;%%
   
\bibitem{Mafra:2019xms}
  C.~R.~Mafra and O.~Schlotterer,
  ``One-loop open-string integrals from differential equations: all-order alpha'-expansions at n points,''
 \eprintN{1908.10830}.
  %%CITATION = ARXIV:1908.10830;%%

\bibitem{Gerken:2019cxz}
  J.~E.~Gerken, A.~Kleinschmidt and O.~Schlotterer,
  ``All-order differential equations for one-loop closed-string integrals and modular graph forms,''
  \eprintN{1911.03476}.
  %%CITATION = ARXIV:1911.03476;%%
 
 
\bibitem{Zerbini:2018sox}
  F.~Zerbini,
  ``Elliptic multiple zeta values, modular graph functions and genus 1 superstring scattering amplitudes,''
  PhD thesis University of Bonn (2018),
  \eprintN{1804.07989}.
  %%CITATION = ARXIV:1804.07989;%% 
 
\bibitem{Gerken:2018jrq}
  J.~E.~Gerken, A.~Kleinschmidt and O.~Schlotterer,
  ``Heterotic-string amplitudes at one loop: modular graph forms and relations to open strings,''
  \doi{JHEP {\bf 1901} (2019) 052}{doi:10.1007/JHEP01(2019)052}
  \eprintN{1811.02548}.
  %%CITATION = doi:10.1007/JHEP01(2019)052;%%
   
\bibitem{Zagier:2019eus}
  D.~Zagier and F.~Zerbini,
  ``Genus-zero and genus-one string amplitudes and special multiple zeta values,''
  arXiv:1906.12339 [math.NT].
  %%CITATION = ARXIV:1906.12339;%%  

\bibitem{Arutyunov:2016etw}
  G.~Arutyunov, D.~Dorigoni and S.~Savin,
  ``Resurgence of the dressing phase for AdS$_{5}\times$ S$^{5}$,''
  \doi{JHEP {\bf 1701} (2017) 055}{doi:10.1007/JHEP01(2017)055}
  \eprintN{1608.03797}.


\bibitem{Dorigoni:2019yoq}
  D.~Dorigoni and A.~Kleinschmidt,
  ``Modular graph functions and asymptotic expansions of Poincar\'e series,''
  Commun.\ Num.\ Theor.\ Phys.\  {\bf 13} (2019) no.3,  569
  \eprintN{1903.09250}.


\bibitem{Kozcaz:2016wvy}
  C.~Koz\c{c}az, T.~Sulejmanpasic, Y.~Tanizaki and M.~\"Unsal,
  ``Cheshire Cat resurgence, Self-resurgence and Quasi-Exact Solvable Systems,''
  \doi{Commun.\ Math.\ Phys.\  {\bf 364} (2018)  835}{doi:10.1007/s00220-018-3281-y}
  \eprintN{1609.06198}.
   
  \bibitem{Dorigoni:2017smz}
  D.~Dorigoni and P.~Glass,
  ``The grin of Cheshire cat resurgence from supersymmetric localization,''
  \doi{SciPost Phys.\  {\bf 4} (2018) 012}{doi:10.21468/SciPostPhys.4.2.012}
  \eprintN{1711.04802}.

\bibitem{Apostol}
T. M. Apostol, ``Modular Functions and Dirichlet Series in Number Theory'', Graduate Texts in Mathematics, Volume 41, Springer.


\bibitem{Bergshoeff:2008qq}
  E.~A.~Bergshoeff, J.~Hartong, A.~Ploegh and D.~Sorokin,
  ``Q-instantons,''
  \doi{JHEP {\bf 0806} (2008) 028}{doi:10.1088/1126-6708/2008/06/028}
  \eprintN{0801.4956}.
  %%CITATION = doi:10.1088/1126-6708/2008/06/028;%%

\bibitem{Hohenegger:2017kqy}
  S.~Hohenegger and S.~Stieberger,
  ``Monodromy Relations in Higher-Loop String Amplitudes,''
  \doi{Nucl.\ Phys.\ B {\bf 925} (2017) 63}{doi:10.1016/j.nuclphysb.2017.09.020}
  \eprintN{1702.04963}.
  %%CITATION = doi:10.1016/j.nuclphysb.2017.09.020;%%

\bibitem{Cheng:2018vpl}
  M.~C.~N.~Cheng, S.~Chun, F.~Ferrari, S.~Gukov and S.~M.~Harrison,
  ``3d Modularity,''
  \doi{JHEP {\bf 1910} (2019) 010}{doi:10.1007/JHEP10(2019)010}
  \eprintN{1809.10148}.
  
  \bibitem{Zagier:1991}
  D.~Zagier,  ``Periods of modular forms and Jacobi theta functions,'' 
  \doi{Invent. Math. {\bf 104}, 449-465 (1991)}{doi:10.1007/BF01245085}.

\bibitem{ZagierFR}
 D.~Zagier, ``Quelques cons\'equences surprenantes de la cohomologie de $SL(2,\mathbb{Z})$,''
  in Le\c{c}ons de Math\'ematiques d'aujourd'hui, Cassini, Paris (2000), 99-123.

\bibitem{Schlotterer:2012ny}
  O.~Schlotterer and S.~Stieberger,
  ``Motivic Multiple Zeta Values and Superstring Amplitudes,''
  \doi{J.\ Phys.\ A {\bf 46} (2013) 475401}{doi:10.1088/1751-8113/46/47/475401}
  \eprintN{1205.1516}.
  %%CITATION = doi:10.1088/1751-8113/46/47/475401;%%

\bibitem{Schnetz:2013hqa}
  O.~Schnetz,
  ``Graphical functions and single-valued multiple polylogarithms,''
  \doi{Commun.\ Num.\ Theor.\ Phys.\  {\bf 08} (2014) 589}{doi:10.4310/CNTP.2014.v8.n4.a1}
  \eprintNT{1302.6445}.
  %%CITATION = doi:10.4310/CNTP.2014.v8.n4.a1;%%

\bibitem{Brown:2013gia}
  F.~Brown,
  ``Single-valued Motivic Periods and Multiple Zeta Values,''
  \doi{SIGMA {\bf 2} (2014) e25}{doi:10.1017/fms.2014.18}
  \eprintNT{1309.5309}.
  %%CITATION = doi:10.1017/fms.2014.18;%%

\bibitem{Stieberger:2013wea}
  S.~Stieberger,
  ``Closed superstring amplitudes, single-valued multiple zeta values and the Deligne associator,''
  \doi{J.\ Phys.\ A {\bf 47} (2014) 155401}{doi:10.1088/1751-8113/47/15/155401}
  \eprintN{1310.3259}.
  %%CITATION = doi:10.1088/1751-8113/47/15/155401;%%  

\bibitem{Stieberger:2014hba}
  S.~Stieberger and T.~R.~Taylor,
  ``Closed String Amplitudes as Single-Valued Open String Amplitudes,''
  \doi{Nucl.\ Phys.\ B {\bf 881} (2014) 269}{doi:10.1016/j.nuclphysb.2014.02.005}
  \eprintN{1401.1218}.
  %%CITATION = doi:10.1016/j.nuclphysb.2014.02.005;%%

\bibitem{Schlotterer:2018zce}
  O.~Schlotterer and O.~Schnetz,
  ``Closed strings as single-valued open strings: A genus-zero derivation,''
  \doi{J.\ Phys.\ A {\bf 52} (2019) no.4,  045401}{doi:10.1088/1751-8121/aaea14}
  \eprintN{1808.00713}.
  %%CITATION = doi:10.1088/1751-8121/aaea14;%%

\bibitem{Brown:2018omk}
  F.~Brown and C.~Dupont,
  ``Single-valued integration and double copy,''
  \eprintNT{1810.07682}.
  %%CITATION = ARXIV:1810.07682;%%

\bibitem{Brown:2019wna}
  F.~Brown and C.~Dupont,
  ``Single-valued integration and superstring amplitudes in genus zero,''
  \eprintNT{1910.01107}.
  %%CITATION = ARXIV:1910.01107;%%
  
  
\end{thebibliography}
\end{document}